# Fragmentation model and strewn field estimation for meteoroids entry

***Limonta S.*[a], Trisolini M.[b], Frey S.[c], Colombo C.[d]***

[a,b,c,d] *Politecnico di Milano, Department of Aerospace Science and Technologies, Via La Masa 34, 20156 Milano*

[a] *simone.limonta@mail.polimi.it*
[b] *mirko.trisolini@polimi.it*
[c] *stefan.frey@polimi.it*
[d] *camilla.colombo@polimi.it*
* *Corresponding Author*

A B S T R A C T

Everyday thousands of meteoroids enter the Earth's atmosphere. The vast majority burn up harmlessly during the descent, but the larger objects survive, occasionally experiencing intense fragmentation events, and reach the ground. These events can pose a non-negligible threat for a village or a small city; therefore, models of asteroid fragmentation, together with accurate post breakup trajectory and strewn field estimation, are needed to enable a reliable risk assessment of these hazards.
In this work, a comprehensive methodology to describe meteoroids entry, fragmentation, descent, and strewn field is presented by means of a continuum approach. At breakup, a modified version of the NASA Standard Breakup Model is used to generate the fragments distribution in terms of their area-to-mass ratio and ejection velocity. This distribution, combined with the meteoroid state, is directly propagated using the continuity equation coupled with the non-linear entry dynamics. At each time step, the fragments probability density time-evolution is reconstructed using Gaussian Mixture Model interpolation. Using this information is then possible to estimate the meteoroid's ground impact probability.
This approach departs from the current state-of-the-art models: it has the flexibility to include large fragmentation events while maintaining a continuum formulation for a better physical representation of the phenomenon. The methodology is also characterised by a modular structure, so that updated asteroids fragmentation models can be readily integrated into the framework, allowing a continuously improving prediction of re-entry and fragmentation events.
The propagation of the fragments' density and its reconstruction, at the moment considering only one fragmentation point, is first compared against Monte Carlo simulations, and then against real observations. Both deceleration due to atmospheric drag and ablation due to aerothermodynamics effects have been considered.

## 1. Introduction

Along its orbit around the Sun, the Earth continuously collects interplanetary dust, rocks, and small grains. Most of the times these rocks are of negligible dimension and burn up harmlessly in the atmosphere. However, larger ones can survive the descent, reaching the ground. It is estimated that more than 1000 kg of meteoroid material reaches the surface of the Earth every day (Passey and Melosh, 1980). These mid-sized asteroids may not be large enough to cause cratering or global scale effects but can still produce significant ground damage and be a real threat for villages or small cities, as demonstrated by the recent Chelyabinsk event (Popova et al., 2013).
To enable the assessment of these risks, models capable of predicting the asteroids fate during entry and fragmentations are needed, together with accurate post breakup trajectory simulations. In literature, few approaches for modelling meteoroid fragmentation exists and are often useful for approximating the fragmentation process and the kinetic energy deposition in the atmosphere. However, a more refined treatment of the breakup and of the interactions between individual fragments is required as base for building reliable and predictive models.
For this reason, in the interest of developing a physically consistent fragmentation model suitable for probabilistic analysis, the present work introduces a continuous, semi-analytical approach, for modelling meteoroids entry events.

### *1.1. Background*

Several models describing the fragmentation of meteoroids under the action of aerodynamic forces can be found in literature. Depending on the approach used, they can be divided into two categories: semi-analytical models and hydrodynamics models (Register et al., 2020, Artemieva and Shuvalov, 2001). In the first category, three distinct families can be further identified: continuous models (Hills and Goda, 1993), discrete models (Andrushchenko and Syzranova, 2019; Artemieva and Shuvalov, 2001; Ceplecha and ReVelle, 2005; Mehta et al., 2015; Douglas O. ReVelle, 2006), and hybrid models (Register et al., 2017; Wheeler et al., 2017). In the following will be given a brief description of each one of them.
Hydrocode simulations consider the object in a quasi-liquid state, evolving in a hypersonic flow. They can capture the detailed flow physics and material properties; however, giver their computational cost, they are not suitable for a probabilistic approach to risk assessment. Some of these models address the interactions between fragments, but are constrained to specific configurations (Register et al., 2020) or with limited number of fragments (Artemieva and Shuvalov, 2001).

On the contrary, semi-analytical breakup models allow probabilistic analyses by introducing simplifying assumptions. They are typically based on the single-body meteor physics equations (Öpik, 1959) and the fragmentation events are assumed to occur when the dynamic pressure at the stagnation point of a bolide exceeds the meteoroid yield strength (Mehta et al., 2015; Passey and Melosh, 1980). The fragmentation products are typically represented either as a cloud-like structure, as a set of discrete fragments, or a combination of both.
The most basic example of continuous model is the so called "pancake model" (Hills and Goda, 1993). At the breakup point, the meteoroid becomes a cloud of continuously fragmenting material. The cloud starts as a sphere and behaves as a single deforming body. During the descent it begins to spread out and flatten due to pressure differences between the front and sides of the debris cloud. While the body is expanding the void, that should form between the small fragments, is instead occupied by other debris continuously created by the fragmentation. This model provides a good description of the energy deposition but does not allow for variations that could result from non-uniform asteroid structures and the behaviour of large, independent fragments.
Examples of discrete fragmentation models are given by the "collective wake model" (Ceplecha and ReVelle, 2005) and "non collective wake model" (D. O. ReVelle, 2006). They both assumes that at the breakup the meteoroid divides in two identical child-fragments, whose strength depends on the parent asteroid strength by means of a Weibull scaling law (Cotto-Figueroa et al., 2016). After few steps, a cloud of identical fragments is produced. The two models differentiate themselves in how the fragments interacts with each other. In the first case, the fragments move side by side and proceed under the same bow shock, increasing the total frontal area and conserving the original object's mass. In the second case, one of the child-fragments is lost to the wake, so that at each fragmentation the area is preserved, and the mass is halved. The assumption of two even fragments resulting from each break is strong, but it could represent the average rate of fragmentation. It should be noted the existence of a geometric inconsistency in the "collective wake model" scheme, as spheres consisting of half the original mass will not double the drag area (Register et al., 2017).
The "independent wake model" (Mehta et al., 2015) follows a more general approach, which considers the two fragments generated at each breakup event to behave independently, and assigns to each of them a lateral spread velocity. The main disadvantage of this scheme is that it does not consider multiple fragmentations at the same time. Moreover, it is assumed that the fragments produced at each breakup will be stronger than their parent fragment, because the breakup would eliminate some of the larger structural weaknesses. However, it is possible that a piece of meteoroid will develop new fractures that could reduce its new strength.
The most recent example of hybrid model combines the features of the "independent wake model" and the "pancake model" to capture both continuous and discrete variation of the kinetic energy of the meteoroids (Register et al., 2017; Wheeler et al., 2017). At the breakup, the bolide is assumed to break into three objects: two spherical fragments and a dust cloud. The cloud is modelled as a continuum using the pancake approximation. At the same time, the two child-fragments continue their descent independently until a new fragmentation point is reached. At each fragmentation event, two new child-fragments and a new dust cloud are formed. This model well reproduces the observed light curves but maintains the limitations of the pancake approach when modelling smaller fragments. The discrete part, instead, considers all the fragments' velocities unchanged with respect to the original body and does not consider a side velocity component so that cannot be used for three-dimensional analyses.
The "sandbag model" (Artemieva and Shuvalov, 2001) is an example of semi-analytical model based on results of detailed hydrodynamics simulations. The deforming meteoroid is represented as a cone rather than a sphere by including both streamwise and spanwise separation. The main disadvantage of the presented model, according to the author, is a strong deficiency of very small fragments (much smaller than the largest one).
Lastly, the "multi-component fragment cloud model" (Newland et al., 2019) assumes that the asteroid has a non-uniform internal structure. The initial body is constructed as a multi-component object comprising different structural groups with different initial strengths. The most fragile components will break off and begin the fragmentation at the highest altitude, while the most resilient components will start to break off closer to the Earth surface. The main criticality of this model is the arbitrariness of the initial composition of the body and the intrinsically discretised nature of the debris cloud: for each group of strength, all the fragments' parameters are identical to each other.
This work presents the development of a methodology for the probabilistic assessment of asteroids re-entry, fragmentation, and impact. The development of the Asteroid Breakup Model (ABM), a continuous approach, which provides a statistical description of the fragments generated at the breakup point, in terms of velocity, flight-path angle, and area-to-mass ratio is described in Sec.2. The dynamics of the problem is presented in Sec.3; in Sec.4, a density-based method, alternative to the traditional Monte Carlo approach, is used for the estimation of the fragment's evolution during the descent. In Sec.5, the methodology is used for the identification of the on-ground footprint of the meteoroid's fragments generated during the descent. This analysis aims at modelling the re-entry and breakup of small asteroids and large meteoroids, hence excluding the rubble pile asteroids. In Sec.6, the methodology is compared against a traditional Monte Carlo approach, while in Sec.7 it has been applied to model a real entry scenario, the 2008TC3 entry.

## 2. Asteroid breakup model

During the atmospheric entry, asteroids can fragment one or multiple times along their trajectory due to the high thermomechanical loads caused by the interaction with the Earth's atmosphere. At each fragmentation point, the body can be either divided in two halves, together with dust and smaller fragments or completely destructed in many small pieces (Ceplecha et al., 1998). The asteroid breakup model presented in this work aims to be a comprehensive representation of this phenomenon. It can describe the properties of the fragments generated after each fragmentation event by means of a probability density function.

The atmospheric fragmentation of a meteoroid is a rarely observed phenomenon. As such, few information is currently available to model meteoroids breakup events. For this reason, we introduce a statistical approach to the phenomenon by borrowing the NASA Standard Breakup Model (SBM) (Johnson et al., 2001; Krisko, 2011) from the space debris field. The NASA SBM uses a probabilistic approach to model in-orbit fragmentation events of spacecraft and rocket bodies due to collisions and explosions. This model is also used for the re-entry breakup analysis in the ESA Debris Risk Assessment and Mitigation Analysis (DRAMA) suite (Gelhaus et al., 2014; R. Kanzler and T. Lips, 2017). Given its flexibility and heritage, in this work, we propose an extension of this model to the asteroid breakup field.

The NASA SBM model describes the fragments generated from an explosion or collision in terms of number, size, and area-to-mass ratio distributions. In addition, it defines the fragments ejection velocity distribution with respect to the parent object. Even if the equations of the NASA SBM model can fully characterise a fragments cloud, its applicability is limited to 1 mm to meter-sized fragments and it does not guarantee mass conservation (Krisko, 2011). Therefore, modifications are required to fully adapt the model to the description of asteroid breakup events.

In the following these modifications are discussed, and the steps required to transform the NASA SBM in the new ABM are outlined. Specifically, Sec.2.1 describes the changes required to the characteristic length distribution, Sec.2.2 to the area-to-mass ratio distribution, Sec.2.3 to the mass conservation, and Sec.2.4 to the velocity distribution. Finally, in Sec.2.5 all the proposed modifications are implemented together and the ABM is obtained. As stated in Sec.1.2, the goal of the ABM is to produce a statistical description of the fragments generated at the breakup point by means of a distribution function. Having its origin from the NASA SBM, this function will be a probability density function (PDF). In this case, is a three-dimensional function in the area-to-mass ratio $(A/M)$, velocity $(v)$ and flight-path angle $(\gamma)$ space. These parameters have been selected because they can appropriately characterise the fragment behaviour in the cloud. The procedure described in the following sections considers a planar case (three-state model) to allow for a clearer description of the model; however, the ABM can be readily extended to a three-dimensional case (six-state model). Sec.2.6 shows the results of the model when applied to the entry of a meteoroid.

### *2.1. Characteristic length distribution*

The empirical characteristic length distribution of the fragments suggested by the NASA SBM is a power law distribution with the following expression:

$$N_c = k\, L_c^{-f} \tag{1}$$

where $N_c$ is the number or generated fragments larger than a given characteristic length $L_c$, $f$ is a fixed scale factor ($f = 1.6$) and $k$ is a tuning parameter that depends on the object that undergoes fragmentation (Johnson et al., 2001).

In many cases in nature, fragmentation results in a fractal distribution: fragments produced by explosions and impacts often shows this behaviour (Turcotte, 1986) that can be mathematically represented by means of a power law, as in Eq. (1). This consideration supports the idea of extending the NASA SBM characteristic length distribution to the breakup of meteoroids, assuming they will exhibit the same behaviour. The validity of such an extension is also supported by the analysis of the meteorites collected on the ground after entry events characterised by intense fragmentations. Although, these findings are biased from the ablation process and from the multiple fragmentations, the number of fragments in relation to their mass approximately follows a power low distribution (Badyukov and Dudorov, 2013; Fries et al., 2014).

Starting from the cumulative distribution of Eq. (1), it is possible to obtain the characteristic length PDF with few steps. Firstly, the total number of fragments produced at the breakup, can be expressed as

$$N_{tot} = k\,(L_{min}^{-f} - L_{max}^{-f}) = k\,\alpha \tag{2}$$

where $L_{min}$ and $L_{max}$ are the minimum and maximum characteristic length of the fragments generated at the breakup, respectively. $\alpha$ is a constant once the characteristic length range is fixed and it is introduced to simplify the expression of the model.

By combining Eqs. (1) and (2) it is possible to obtain the Cumulative Distribution Function (CDF) of the characteristic length:

$$CDF_L = \frac{N_{tot} - N_c}{N_{tot}} \tag{3}$$

At this point the characteristic length PDF can be derived by taking the derivative of the $CDF_L$ with respect to $L_c$ as follows:

$$p_L = \frac{dCDF_L}{dL_c} = \frac{f}{\alpha}\, L_c^{-f-1} \tag{4}$$

### *2.2. Area-to-Mass distribution*

In the NASA SBM, the area-to-mass $(A/M)$ distribution assumes that the fragments generated at breakup have a relatively wide range of density (Johnson et al., 2001). This assumption is reasonable for spacecrafts or rocket bodies because they are composed of different materials, but it cannot be applied to meteoroids without a complete change in the distribution shape. In this analysis, we assume meteoroids have uniform density so that the original $A/M$ distribution must be modified.

To conserve the fragments density, we assume the bodies have a spherical shape. This is a strong assumption; however, it is a common choice in most of the models in literature (Mehta et al., 2015; Register et al., 2017). In fact, the shapes of asteroids are typically irregular and not well known a priori. Additionally, they can be modified by ablation during entry. The spherical shape approximation is therefore considered a suitable compromise between model simplicity and accuracy.

As direct consequence of this approximation, the characteristic length is defined as the object diameter and the relations between the fragments geometrical parameters can be expressed as follows:

$$L_c = 2\,R \tag{5}$$

$$A/M = \frac{3}{2\,\rho_m\,L_c} \tag{6}$$

$$M = \left(\frac{3}{4\,\rho_m}\right)^2 \frac{\pi}{(A/M)^3} \tag{7}$$

where $\rho_m$ is the meteoroid density, $M$ is the meteoroid mass, and $R$ is the radius. By means of Eq. (6), together with Eq. (4), it is possible to obtain an alternative expression for $p_L$ expressed in terms of $A/M$ by exploiting the coordinate transformation outlined in Appendix A. The PDF distribution obtained is the following:

$$p_{A/M} = \frac{f}{\alpha}\left(\frac{2}{3}\rho_m\right)^f\ A/M^{f-1} \tag{8}$$

It represents the number of fragments (normalised) in the interval $[A/M, A/M + d(A/M)]$. As expected, looking at the representation of the function (Fig. 1), higher probability value is given to smaller fragments (larger $A/M$ value).

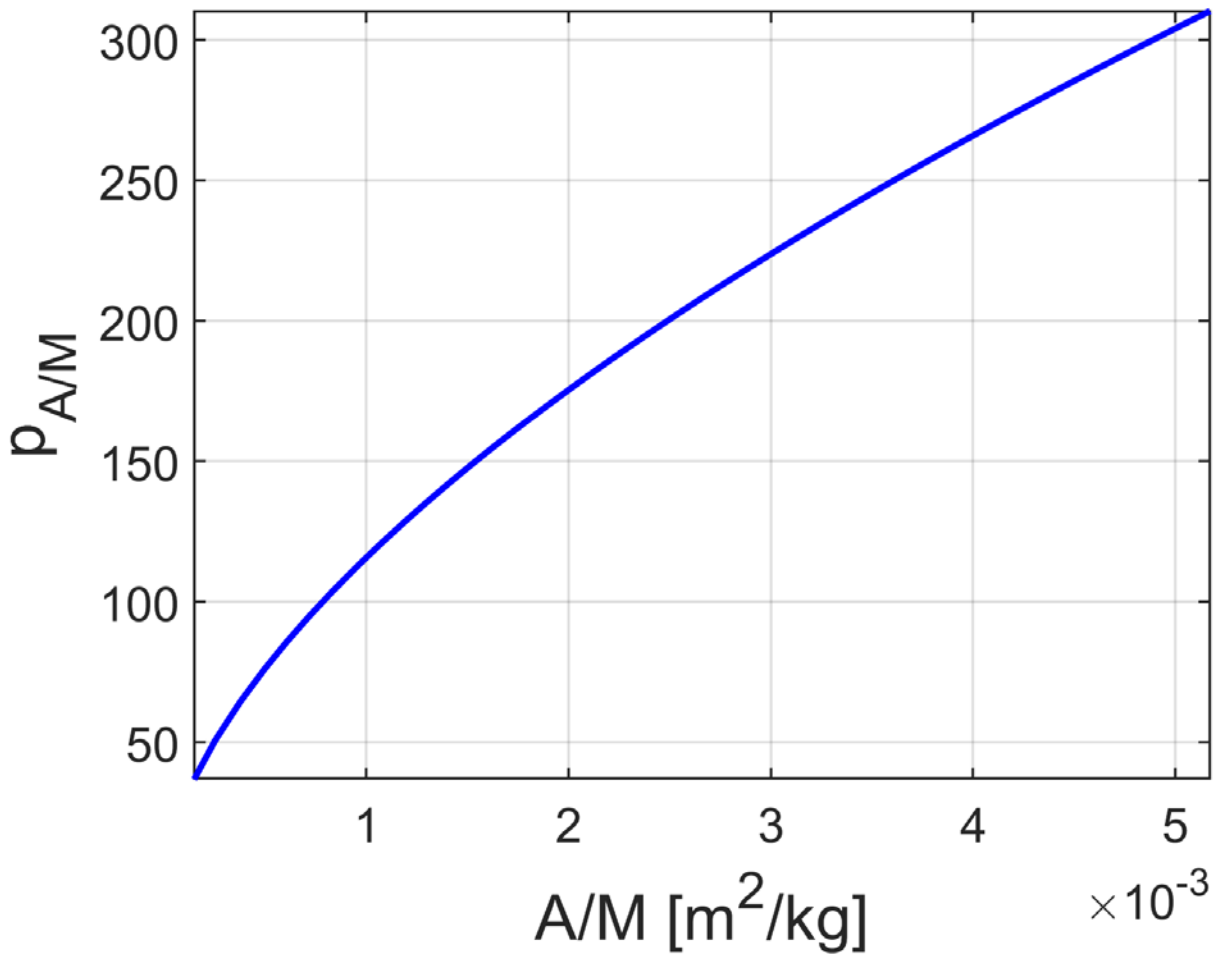

Fig. 1. $p_{A/M}$ distribution for a meteoroid with $L_c$ = 5 m considering fragments size between 0.1 m and 3.5 m. (ρ = 2900 kg/m[3]).

### *2.3. Mass conservation*

In literature no univocal method is prescribed to implement the mass conservation in the NASA SBM (Krisko, 2011). For example, it is possible to use an iterative scheme, suitable for a Monte Carlo approach (Bade et al., 2000), to generate fragments until the total fragmentation mass is modelled. However, for the purpose of this work a distribution function that guarantees the mass conservation is required.

It is possible to exploit the power law distribution of Eq. (1) to impose the mass conservation. The tuning parameter $k$ depends on the object that undergoes the fragmentation process and can be used to impose the mass conservation. First, we rewrite the probability density function in terms of the fragments mass Eq. (7). By exploiting the same approach of Sec.2.2 and outlined in Appendix A , Eq. (8) is transformed into:

$$p_M = \frac{f}{3\alpha}\left(\frac{\pi}{6}\,\rho_m\right)^{f/3}\ M^{-f/3-1} \tag{9}$$

$p_M$ is the result of a coordinate transformation, hence it still represents the frequency of a fragment of a given mass (i.e. $p_M$ represents the normalised number of fragments having mass inside the interval $[M_i, M_i + dM_i]$), therefore the following properties are still valid:

$$\int_{M_{min}}^{M_{max}} p_M\,dM = 1 \tag{10}$$

and

$$\int_{M_{min}}^{M_{max}} p_M \, N_{tot} \, dM = N_{tot} \tag{11}$$

To obtain the normalised mass of the fragments it is sufficient to multiply $p_M$ (i.e. the normalised number of fragments) by $M$, as in Eq. (11). The total mass is found by multiplying the result by the total number of fragments, $N_{tot}$, as shown in Eq. (12). The parameter $k$ is then found by imposing the mass conservation as follows:

$$N_{tot} \int_{M_{min}}^{M_{max}} M \, p_M \, dM \; = M_{tot} \tag{12}$$

Whose solution is:

$$k = \frac{M_{tot}}{\frac{f}{3-f}\left(\frac{\pi}{6}\,\rho_m\right)^{f/3}\left(M_{max}^{1-f/3} - M_{min}^{1-f/3}\right)} \tag{13}$$

Eq.(13) shows that the parameter $k$ can be determined by defining the mass range of the generated fragments. This condition is translated in a boundary definition of the minimum and maximum characteristic length of the fragments in the cloud. However, there is not a general rule to select them, and they must be selected depending on the specific case in exam (Appendix B).

### *2.4. Velocity distribution*

In the NASA SBM, the velocity distribution is modelled as a log-normal PDF as follows:

$$p_{(\nu|\chi)} = \mathcal{N}(\mu, \sigma) \tag{14}$$

where $\mathcal{N}$ is the normal Gaussian distribution defined by the following parameters:

$$\begin{aligned} \sigma &= 0.4 \\ \mu &= 0.2\,\chi + 1.85 \\ \chi &= Log_{10}(A/M) \\ \nu &= Log_{10}(\Delta v) \end{aligned} \tag{15}$$

However, this model only expresses the magnitude of the velocity variation, not its direction. A model of the ejection direction is then required to properly describe the breakup. Looking at the observations there is no evidence of a preferred ejection direction during fragmentations, so the impulse direction is assumed to be uniformly distributed. To introduce a directional component in the probability density functions of $\Delta v$, the distribution needs to be divided by the surface area that the tip of the velocity vector draws out as explained by Frey et al. (Frey and Colombo, 2020). The resulting distribution is the following:

$$p_{\Delta v} = \frac{p_{\Delta v}}{\mathcal{S}} \tag{16}$$

$$\mathcal{S} = \begin{cases} 2\,\pi\,\Delta v & \text{for planar case} \\ 4\,\pi\,\Delta v^2 & \text{for spherical case} \end{cases} \tag{17}$$

### *2.5. Joint-PDF*

The joint-PDF is the core of the ABM. It is a multidimensional density function that describes the characteristics of the fragments generated by a fragmentation event. Specifically, as stated in Sec.2 it is defined in terms of $A/M, \nu, \gamma$ for the planar case. To obtain this joint distribution, it is sufficient to combine the distributions along each dimension (or state space).

The first step is merging the area-to-mass distribution with the delta velocity distribution. Both PDFs must be projected onto the same phase space, in this case the logarithmic space $(\chi, \nu)$ has been selected. A variables transformation (Appendix A. ) is applied to Eq. (8) obtaining the following PDF in the logarithmic space:

$$p_\chi = \ln(10)\,10^\chi\,p_{A/M} \tag{18}$$

Then the PDF in the (χ, ν) space is obtained by multiplying the two probability density functions (14) and (18):

$$p_{\nu,\chi} = p_{\nu|\chi}\, p_\chi \tag{19}$$

Eq. (19) is then transformed (Appendix A. ) in the $(A/M, \Delta v)$ space (Fig. 2):

$$p_{A/M,\Delta v} = p_{\nu,\chi} \frac{1}{ln^2(10)A/M\ \Delta v} \tag{20}$$

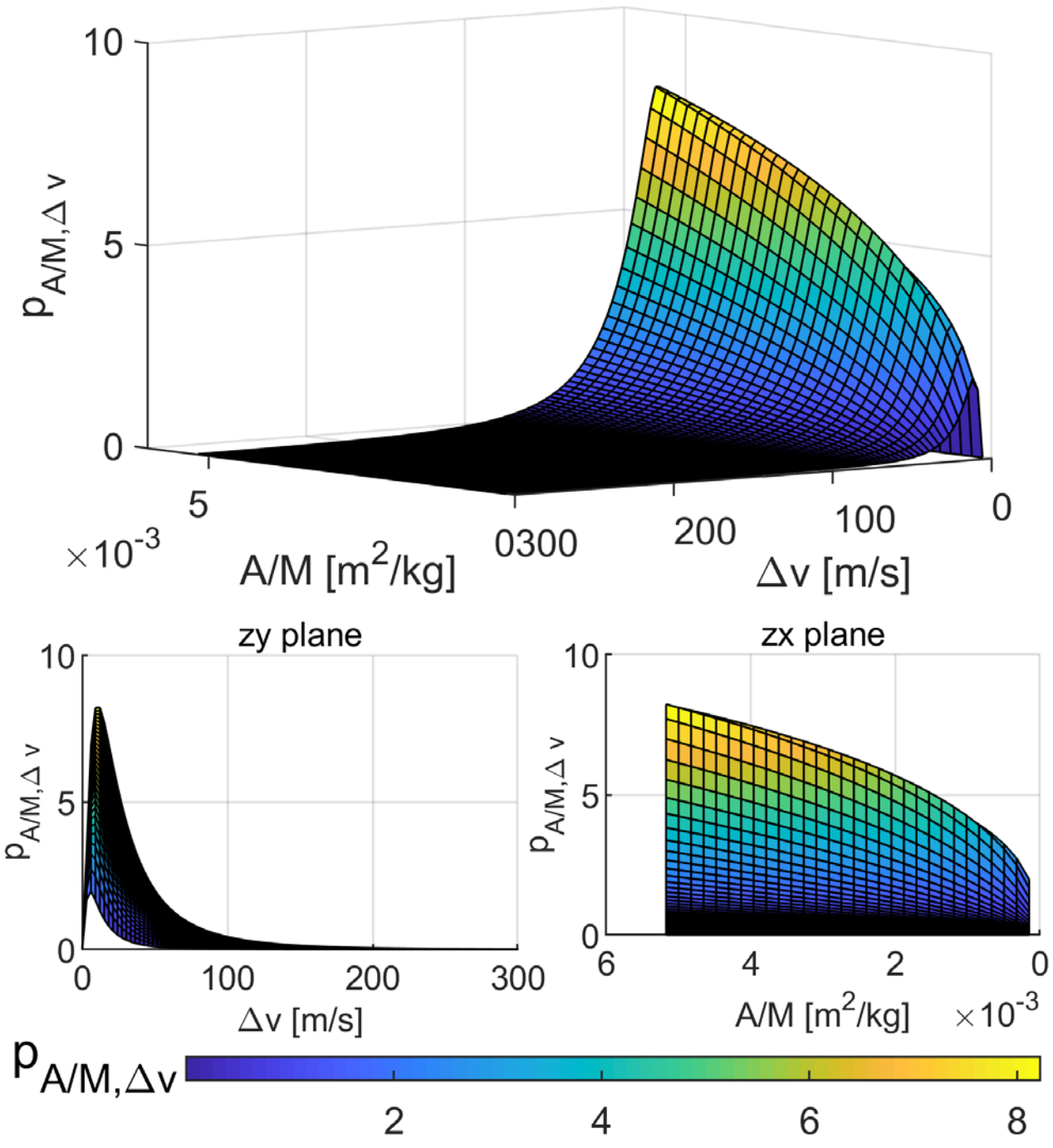

Fig. 2. $p_{A/M,\Delta v}$ distribution for a meteoroid with $L_c$ = 5 m and ρ = 2900 kg/m³ considering fragments size between 0.1 m and 3.5 m.

From Fig. 2 it can be observed that, as expected, the probability value is higher as the fragments size is smaller, and that, on average, a larger velocity increment is given to the smaller fragments. Moreover, the velocity increment range becomes narrower as the fragments becomes larger. As a result, large and heavy fragments tend not to significantly change their trajectory after the breakup. The joint-PDF is then obtained by adding the directional dependence of the impulse as explained in Sec.2.4.

$$p_{A/M,v_x,v_y} = p_{A/M,\Delta v} \frac{1}{\mathcal{S}_{2D}} \tag{21}$$

Where $\mathcal{S}_{2D}$ is by the surface area that the tip of the velocity vector draws out (Eq. (17)) and the subscript 2D stand for the planar case.

The joint PDF is now function of $A/M, v_x, v_y$ where $v_x$ and $v_y$ are the $\Delta v$ component along the trajectory and normal to it at the breakup point. Again, a variable transformation from $v_x, v_y$ to $v, \gamma$ is performed (Appendix A. ) to project the joint-PDF in the desired state space (Fig. 3) as follows:

$$p_{A/M,v,\gamma} = p_{A/M,v_x,v_y}\, |v| \tag{22}$$

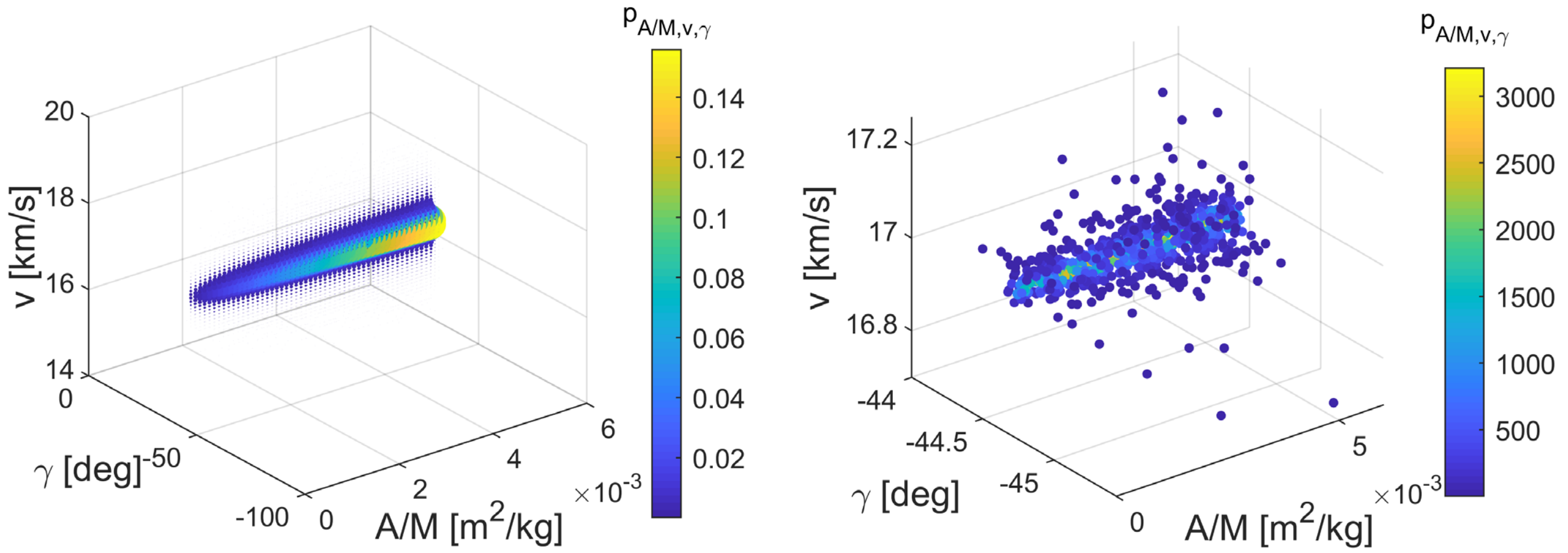

(a) Joint PDF represented with 3D scatter using an uniform spaced grid. The scatter size and colour depends on the density value

(b) Sampling of the joint-PDF.

Fig. 3. Joint PDF. Represented for a meteoroid with $L_c$ = 5 m, v = 17 km/s, γ = 45° and ρ = 2900 kg/m$^3$ and considering characteristic length bounds between 0.1 m and 3.5 m.

Similarly to the distribution of Fig. 2, also in this case a bigger velocity increment is given to the smaller fragments. However, from Fig. 3 it is possible to directly appreciate the distribution in the re-entry state space ($v$, $\gamma$, $\rho$) of the fragments after the breakup instead of their variation. In addition, there is also information about the flight-path angle: as expected, due to the isotropic distribution of $\Delta v$ the lighter the fragments are, the more scattering they receive after the breakup.

### *2.6. Asteroid Breakup Model implementation*

During the atmospheric descent, the fragmentation event is triggered when the dynamic pressure at the stagnation point of the object reaches the meteoroid strength limit as shown in Eq. (23) and it is assumed instantaneous (Mehta et al., 2015; Register et al., 2017). However, it is often found in real cases that the value of the dynamic pressure as computed from real data is considerably lower than the expected and also measured meteoroid yield strength (Devillepoix et al., 2019; Popova et al., 2011).

$$St = \rho\, v^2 \qquad (23)$$

where $St$ is the meteoroid strength limit, $\rho$ is the air density and $v$ the meteoroid velocity at the breakup. At this point the states of the meteoroid at the fragmentation instant are used by the ABM to generate a suitable joint PDF. The distribution of Eq. (22) is then sampled, and a family of fragments to be propagated is generated. The propagation approach will be described in detail in Sec. 5.1. The model has been implemented in MATLAB, version R2018b.

As an example, consider a typical meteoroid entry scenario (described in detail in Sec. 6), with entry velocity $v$ = 17 km/s and entry flight-path angle $\gamma$ = -45°. The sampling of the joint PDF results in a "columnar shape" domain, as shown in Fig. 3. The selected meteoroid has an initial radius of 5 m and density $\rho_m$ = 2900 kg/m$^3$. The ABM requires the definition of the range of the fragments size considered. There is not a fixed rule for the definition of the characteristic length boundaries; they must be selected depending on the analysis objective and data available. Further details will be given in Sec. 6.

As Fig. 3b shows, the $\Delta v$ generated by the ABM does not create a relatively large variation in the fragments' velocity. On the other hand, the fragmentation causes a relatively large variation in the $A/M$ dimension with a predominance of small fragments.

## 3. Dynamic model

The meteoroid is modelled as a point mass with uniform density and given area-to-mass ratio, subject to Earth gravity, air resistance and ablation. Lift and sides forces, accordingly to the most recent models in literature (Mehta et al., 2015; Register et al., 2017), are neglected, as meteoroids are in general heavy and non-aerodynamic bodies. Moreover, the unknown and variable shape of these objects does not allow to determine reliable values for the lift coefficient.

### *3.1. Planar approach*

Most of the dynamic models used in meteor science are based on a two-dimensional motion. This approach can be justified observing the typical distribution of the fragments on ground: the fragments distribute across an elongated elliptic area, called strewn filed (Norton, 2009). The major axis coincides with the direction of motion of the meteorite swarm and is typically much bigger with respect to the minor axis. Therefore, the bidimensional ground footprint can be approximated as a one-dimensional line in a preliminary analysis.

Following the procedure outlined by Register (Register et al., 2017), the phenomenon is described assuming a planar reference frame over a circular, non-rotating, Earth (Fig. 4). The system of the governing equation can be written as:

$$\begin{cases} \dot{h} = v \sin\gamma \\ \dot{\zeta} = \dfrac{v\cos\gamma}{R_E + h} \\ \dot{v} = \dfrac{\rho\, v^2 A/M\, c_d}{2} - g\sin\gamma \\ \dot{\gamma} = \cos\gamma \left(\dfrac{v}{R_E + h} - \dfrac{g}{v}\right) \\ \dot{A/M} = \dfrac{1}{6}\,\rho\, c_d\, \sigma_{ab}\, (A/M)^2\, v^3 \end{cases} \tag{24}$$

where $h$ is the altitude from the ground, $v$ is the fragment velocity, and $\gamma$ is the flight-path angle. The gravity acceleration ($g$) is modelled as a function of the altitude by means of an inverse square model (Eq. (25)), while for the atmospheric density ($\rho$) an exponential model has been adopted (Eq. (26)). The drag coefficient ($c_d$), can be considered constant at high hypersonic regime, since it is independent on the Mach number. For the case in exam we consider $c_d = 1$ as reference value (Register et al., 2017). $\sigma_{ab}$ is the ablation coefficient. As pointed out by Wheeler (Wheeler et al., 2017), the ablation rate should vary with fragment size, shape, speed, and altitude throughout entry. However, appropriate values for those rates and how much they may vary throughout the entry remain uncertain. In this analysis, we select a constant value of $\sigma_{ab} = 10^{-8}\, s^2 m^{-2}$ (Register et al., 2017; Hills and Goda, 1993), that is defined for the evolution of a fragments cloud. $\zeta$ is an angular variable representing the angular range distance from the atmospheric entry point. The ground distance covered by the flying object is called range and can be found by multiplying $\zeta$ by the Earth radius (Eq. (27)).

$$g = g_0 \frac{G\, M_E}{(R_E + h)^2} \tag{25}$$

$$\rho = \rho_0 \exp\left(\frac{H_2 - h}{H_1}\right) \tag{26}$$

$$\text{range} = \zeta\, R_E \tag{27}$$

where $\rho_0$ is a reference atmospheric density, $H_1$ and $H_2$are constants related to the atmosphere of the planet, $M_E$ is Earth's mass, $R_E$ is Earth's radius and $G$ is the universal gravitational constant ($G$ = 6.67 10−11 $m^3 kg^{-1} s^{-2}$)

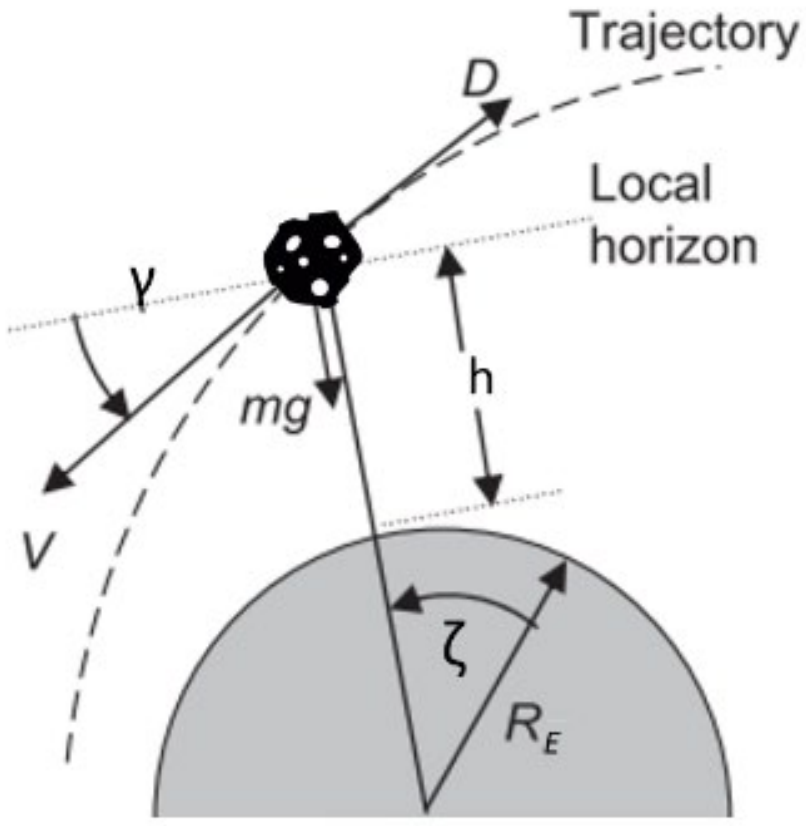

Fig. 4. Planar reference frame (Sforza, 2016).

### *3.2. Three-dimensional extension*

Analysing the meteoroid entry, it is possible to extend the results of the planar dynamics model to describe a full three-dimensional motion. Considering the three-dimensional set of equations governing the descent of a non-lifting object in the atmosphere for a non-rotating Earth (Fig. 5) (Avanzini, 2009):

$$
\begin{cases}
\dot{h} = v \sin\gamma \\
\dot{\lambda} = \dfrac{v \cos\gamma \ \sin\chi}{(R_E + h)\cos\delta} \\
\dot{\delta} = \dfrac{v \cos\gamma \ \cos\chi}{R_E + h} \\
\dot{v} = \dfrac{\rho\, v^2 A/M\, c_d}{2} - \ g \sin\gamma \\
\dot{\chi} = \dfrac{v \cos\gamma \ \sin\chi \tan\delta}{(R_E + h)} \\
\dot{\gamma} = \cos\gamma \left(\dfrac{v}{R_E + h} - \dfrac{g}{v}\right) \\
A\dot{/}M = \dfrac{1}{6}\, \rho\, c_d\, \sigma_{ab}\, (A/M)^2\, v^3
\end{cases}
\tag{28}
$$

The main differences with respect to Eqs. (24) are the presence of the heading angle ($\chi$) defined as the angle between the local meridian and the projection of the velocity vector on the local horizon, and of the latitude ($\delta$) and longitude ($\lambda$) instead of the angular range ($\zeta$). As pointed out by Avanzini (Avanzini, 2009) for a motion constrained over any plane containing a great circle, the equations of motion reduce to the two-dimensional set of Eq. (24). For this reason, the meteoroid entry can be modelled using the two-dimensional equations until the breakup. This approximation, in general, it is not valid when propagating the fragments cloud after the breakup. During fragmentation, in fact, in a three-dimensional analysis, the velocity is scattered also in the out-of-plane direction.

In this case, the heading angle $\chi$ is added as new state in the joint PDF. However, the heading angle variation, during the fragments' descent, can be neglected at small latitude angles without introducing significant error. Following these considerations, $\chi$ can be approximated constant and this allows to decouple the heading angle evolution from the other equations. At this point, the only remaining variables depending on $\chi$ are the latitude $\delta$ and the longitude $\lambda$. Given that $\chi$ is assumed constant in time and that $\delta$ is small (i.e. $\cos\delta \sim 1$), latitude and longitude can be considered as the projection of the angular range ($\zeta$) in two orthogonal direction (North and East). The three-dimensional dynamics reduces to the usual two-dimensional dynamics, with the inclusion, ex-post, of the $\chi$ dependence, used to decompose the angular range in latitude and longitude.

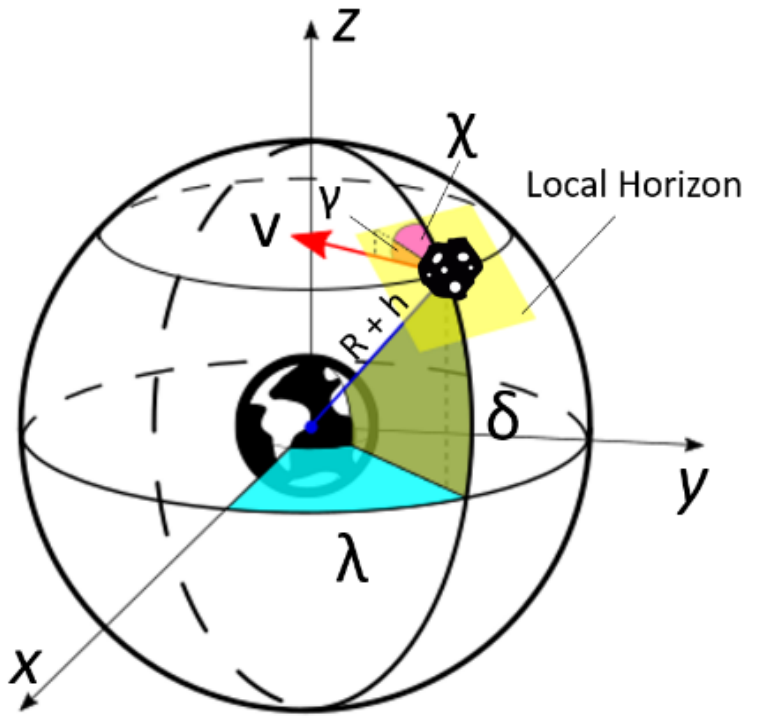

Fig. 5. Spherical reference frame.

Form these considerations it can be derived a method to recover the joint PDF in the three-dimensional physical domain. The $\chi$ angle probability distribution can be considered constant and independent with respect to the planar joint PDF. Specifically, this relation is valid at each time step:

$$
p_{A/M,v,\gamma,\chi} \ = \ p_{A/M,v,\gamma}\ p_\chi \tag{29}
$$

Instead of propagating the full PDF, it is then convenient to propagate only its planar counterpart, consider $p_\chi$ as independent and recover the third-dimension information exploiting the augmented PDF using Eq. (29).

## 4. Density Based Approach

The ABM describes the fragments cloud as a whole, by means of a joint PDF. Its intrinsic continuous nature allows to exploit a novel strategy for the propagation of the fragments' dynamics. This method is based on a continuum approach and, in this paper, it is proposed as alternative to the traditional Monte Carlo methodology.

### 4.1. Probability Density Function evolution

Presented for the first time by Heard (Heard, 1976), the central idea of this approach is to consider the fragments population as a fluid with continuous properties. The coupling between the dynamics and the continuity equation (Eq. (30)) enables the exact evaluation of the evolution in time of the joint PDF. In this way, the analysis of the single objects is abandoned, and the ensemble of fragments together with their density is considered.

Once the initial distribution of the fragments is known, the continuity equation is used to obtain its evolution in time as follows:

$$\frac{\partial n}{\partial t} + \nabla \cdot \boldsymbol{f} = n^{+} - n^{-} \tag{30}$$

where $n$ represents the fragments density in the space of the entry coordinates and the vector field $\boldsymbol{f}$ describe the differential problem. The divergence of $\boldsymbol{f}$ accounts for continuous phenomena (e.g. drag, ablation) and $n^{+}, n^{-}$ are respectively the source and sink terms that model discontinuous events.

This method is quite general and it has also been applied to describe the evolution of interplanetary dust (Heard, 1976), nano-satellites constellations (McInnes, 2000) and debris cloud evolution (Frey et al., 2019a; Letizia, 2016). Trisolini (Trisolini and Colombo, 2021, 2020, 2019) and Halder (Halder and Bhattacharya, 2011) recently applied this approach to the re-entry of spacecraft and their uncertainties. The objective of this work is to extend this methodology to meteoroids entry by introducing the physics related to ablation and fragmentation phenomena.

Starting from Eq. (30) and exploiting the method of characteristics (Halder and Bhattacharya, 2011; McInnes, 2000; Trisolini and Colombo, 2019), together with the entry dynamics of Eq. (24), we obtain the differential equation that describes the joint PDF evolution in time ($\dot{n}$) and that can be propagated with the other differential equations:

$$\dot{n} = \left[\sin\gamma\left(\frac{v}{R_E + h} - \frac{g}{v}\right) + \rho\frac{vA}{Mc_d} - \frac{1}{3}\,\rho\, c_d\, \sigma_{ab}\, A/M\, v^3\right] n \tag{31}$$

The expression of Eq. (31) refers to the planar entry dynamics case. However, the same procedure can be exploited to obtain a three-dimensional density evolution.

### 4.2. Augmented probability density functions

In the formulation of the problem described in Sec.4.1, the density function depends only to the fragment cloud distribution. However, the density-based method allows the extension of the density function. In particular, uncertainties can be modelled as density functions and can thus be added to the three-dimensional joint PDF in ($A/M$, v, γ) to obtain an augmented probability density function. During the meteoroid entry, the main uncertainty source, is considered to be the estimation of the position of the meteoroid in the sky. It is also difficult to predict when, along the trajectory, the fragments generated stop influencing each-other. This time uncertainty can also be considered as a position uncertainty of the meteoroid at the breakup. The position uncertainty is traduced into two independent uncertainties: range and altitude. The uncertainties distributions are modelled as Gaussian normal distribution and are assumed independent from all the other states. Following this procedure, the joint PDF is transformed to a five-dimensional function, and it can be expressed as follows:

$$p_{augmented} = p_{A/M,v,\gamma}\, p_h\, p_{range} \tag{32}$$

where $p_h$ and $p_{range}$ are defined as Gaussian uncertainties. In Fig. 6 the ABM joint PDF has been augmented considering an altitude and range uncertainties having a 10% relative variance. The new augmented joint PDF has been sampled and represented using histograms along each dimension.

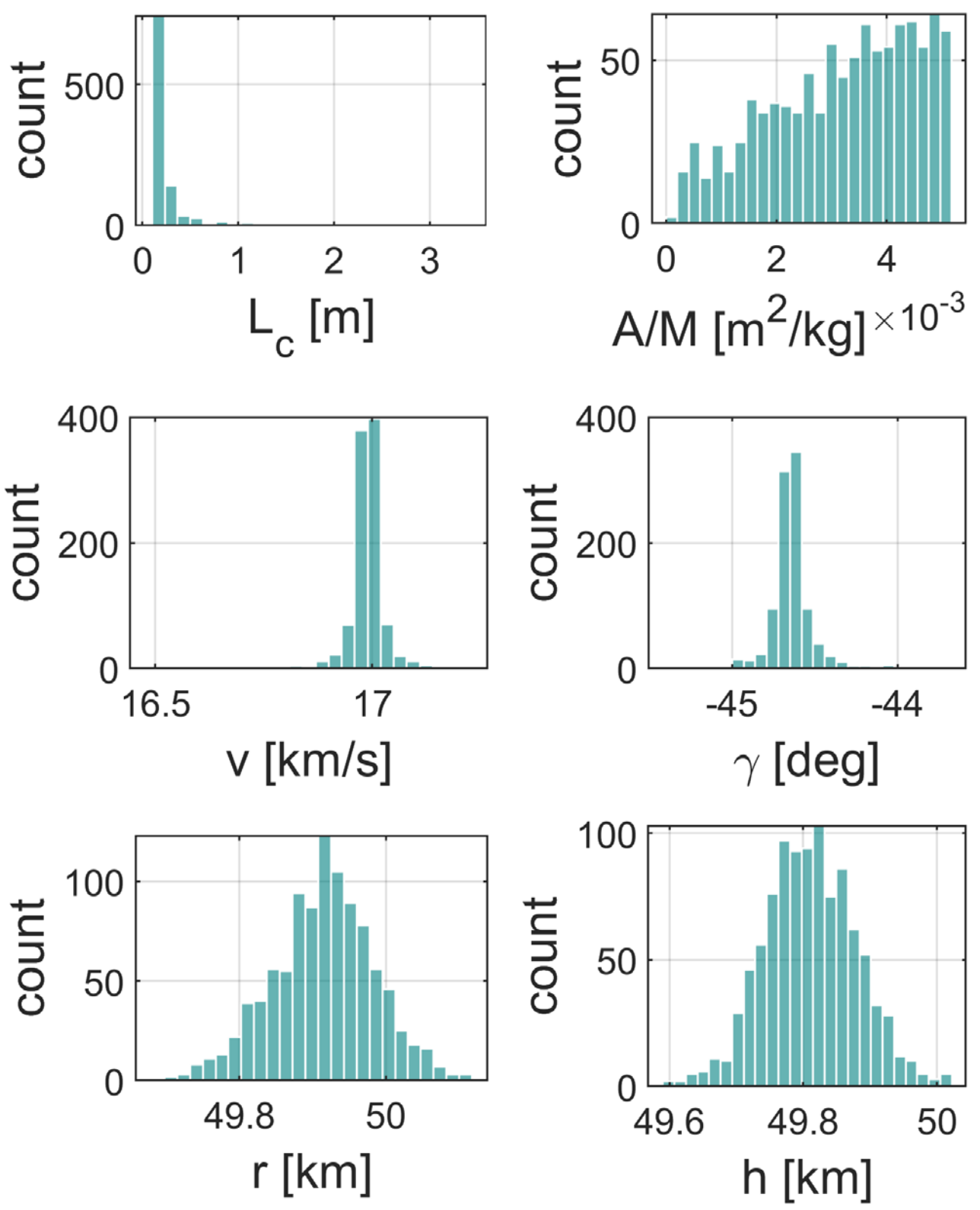

Fig. 6. Histograms of $p_{augmented}$ along each dimension. Count is the number of fragments in each bin.

## 5. Propagation, fitting, and strewn field

This chapter describes the approach used to propagate the fragments generated via the ABM (Sec. 2) until they reach the ground or demise, and to reconstruct the density information at each time step. In addition, two different strategies that simplify the domain and improve the quality of the results are presented. Finally, a procedure to obtain the strewn field distribution starting from the PDF reconstructed is presented.

### *5.1. Fragments cloud propagation*

When the meteoroid enters the atmosphere, it is modelled using the two-dimensional single body dynamics (Sec.3.1) until the breakup. As stated in Sec. 2.5 there is no a general rule for choosing the characteristic length bounds. In this analysis, for a comparative analysis with the MC approach, it is suggested to consider fragments ranging from $L_c = 0.1\ m$ to the 70% of the meteoroid diameter at breakup.

Once a pre-defined number of samples have been generated (Sec. 2.5), the cloud density evolution is simulated exploiting the continuity equation coupled with the dynamics (Sec.4.1) until they reach the ground. The integration is stopped in advance if the fragments either ablate or reach a negligible level of kinetic energy (15 J). It is also assumed that after the entire meteoroid undergoes fragmentation and no other breakup events will occur.

Fig. 7 shows the domain evolution in time highlighting different time instants with different colours, for the same meteoroid used in Sec.2.6. As the time passes, the domain extends along the $A/M$ dimension and shrinks in the other dimensions, progressively transforming into a line. A domain exhibiting this behaviour is challenging to fit, as outlined in Sec.5.2.

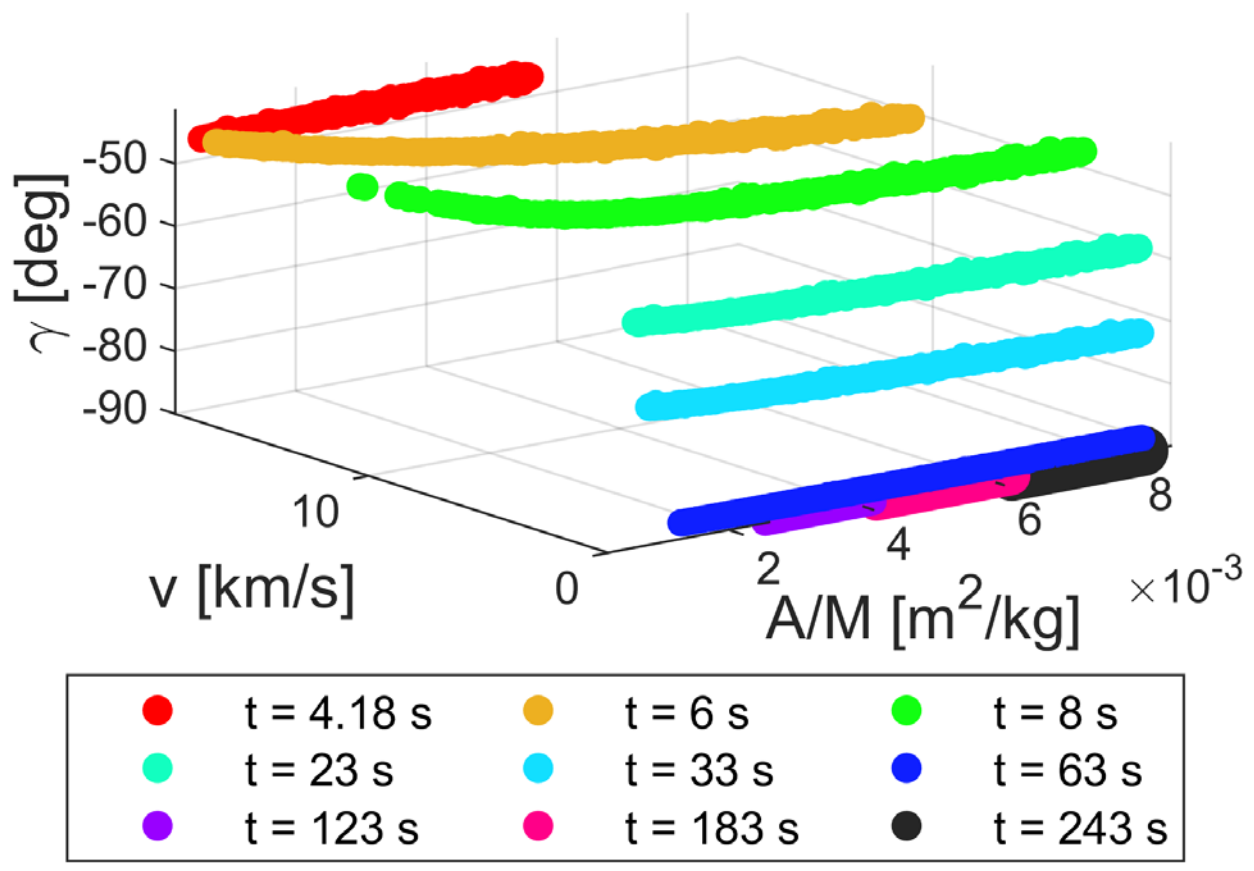

Fig. 7. Representation of the domain evolution in time. The different colours represent different time instants.

### *5.2. Fitting and Marginalisation*

The density-based propagation performed via the method of characteristics results in a discrete set of samples carrying the information of the actual fragment density for the defined state space. To obtain the density information in the entire domain, the distribution must be reconstructed at each integration step by interpolating the scattered data over the state space domain. In this work, the density distribution is reconstructed by fitting it to a Gaussian Mixture Model (GMM). This method has been proposed by Frey et al., 2019b for the reconstruction of the fragments density following a catastrophic fragmentation of a satellite in orbit. The GMM is fit using a gradient descent optimization method to minimize a given cost function dependent on the densities of each sample. This procedure is automatically implemented in the Starling suite, a novel tool developed at the Politecnico di Milano and funded by the COMPASS European Research Council project and the European Space Agency (Frey et al., 2019a). The suite has been designed to estimate evolving continua subject to non-linear dynamics and consists of several independent routines. In this paper, the fitting routine has been exploited, with minor changes to adapt it to the meteoroid entry dynamics. Further details about the Starling suite and its fitting optimisation technique can be found in Frey et al., 2019a, 2019b.

Once the joint PDF has been evaluated, it is possible to compute its marginal along each dimension (i.e., the probability of an event irrespective of the outcome of other variables). For example, the one-dimensional marginal probability along the x-direction of a three-dimensional distribution function $p(x, y, z)$ is expressed by:

$$m_x = \iint p(x, y, z)\, dy\, dz \tag{33}$$

In this framework, the marginalisation of the multivariate normal distribution over one or more distributions is another multivariate normal distribution. The new mean and covariance are simply the partitioned mean and covariance of the marginalised distribution. The extension to the marginalisation of a GMM is trivial.

During the propagation both the density and the volume of the state space deform (Fig. 7). Considering the meteoroid entry scenario, the domain shape increases its complexity over time. In these cases, the fitting routine might not converge and provide inaccurate results. In particular, the samples fitting performed exploiting the Starling suite provides good results only for the initial part of the trajectory, then some simplifications are required to provide more precise results. The main criticality in the domain shape is due to the large $A/M$ range considered. For this reason, in this paper two different strategies to mitigate this problem are discussed: the reduction of the $A/M$ domain and the simplification of the $A/M$ distribution. The first method limits the analysis to the larger fragments in the distribution because they are the ones of main concern for a risk analysis. The second method exploits a binning technique to discretise the $A/M$ dimension and uses interpolation to recover the fragments distribution on the whole domain. In the following the two approaches are described in more detail.

### *5.3. Reduced domain approximation*

With this method, the fragment cloud fitting is performed on a reduced domain along the A/M dimension. This strategy considers only the larger fragments generated at the breakup. The reason behind this choice is to reduce the domain tendency to the elongation: the heavier objects evolve relatively slowly, hence reaching the ground before the shape of the domain becomes too complex. A 1 m diameter threshold has been selected. This choice is an indication, not a general rule, as the selection of the threshold can depend on the initial entry conditions of the meteoroid.

The obtained domain evolution is presented in Fig. 8. Also in this case the shape of the domain is complex, but differently from Fig. 7, it remains relatively confined. With respect to the previous case, the domain range is sensibly reduced for all the variables, thus reducing the criticalities in the fitting process and allowing Starling to reach good fitting results for the whole trajectory. With this strategy the joint PDF and its marginals can be available at every time step.

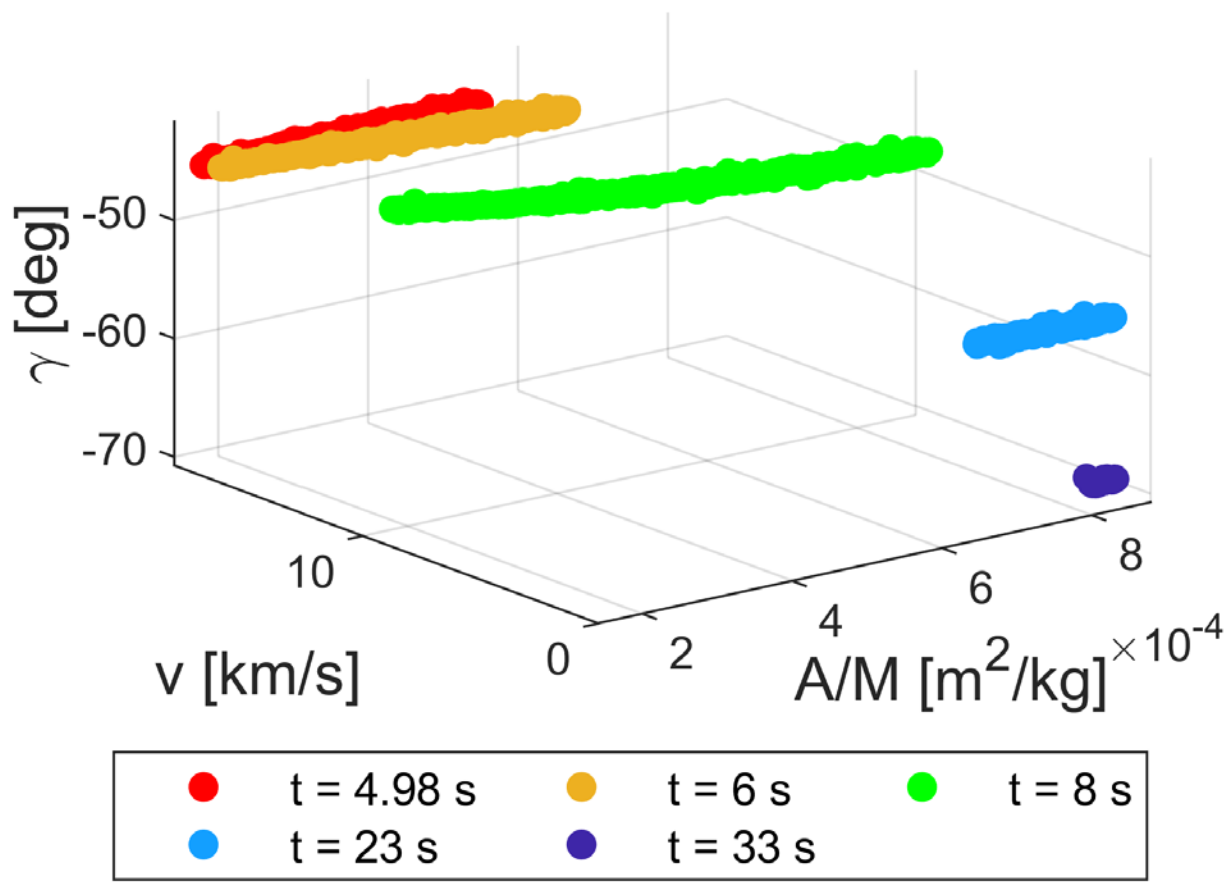

Fig. 8. Domain evolution in time considering only fragments larger than 1 m in diameter.

### *5.4. Area-to-mass binning approximation*

The binning method relies on the approximation of the $A/M$ distribution, where the PDF is approximated using a piecewise linear function. For each bin, a representative object is propagated. At the end, the fitted PDF of each representative object are summed together and weighted according to the probability density associated to each bin. A number $N_b$ of bins in area-to-mass ratio is defined, for each bin an average $A/M$ is assumed and the corresponding partial density is obtained according to the $A/M$ probability density function (Eq.(8)). For a preliminary analysis $N_b = 10$ bins are selected and, for each one of them, a representative object is chosen as the average between the bin edges:

$$A/M_{bin_i} = 1/2\ (A/M_i + A/M_{i+1}) \tag{34}$$

The probability density considered is then the probability of each $A/M$ bin multiplied by the weight of each bin over the domain (i.e. the area associated with each bin).

$$w = \int_{A/M_i}^{A/M_{i+1}} PDF_{A/M}\, dA/M \tag{35}$$

$$PDF_{A/M_{bin_i}} = w\, PDF_{A/M}(A/M_{bin_i}) \tag{36}$$

In this way the conservation of the area under the function is guaranteed. Different binning strategies can be adopted. After comparing different options (Limonta et al., 2020), the log-spaced strategy has been selected. Compared to a uniform binning, the log-spaced option has a denser discretisation for the lighter fragments without neglecting the heavier ones.

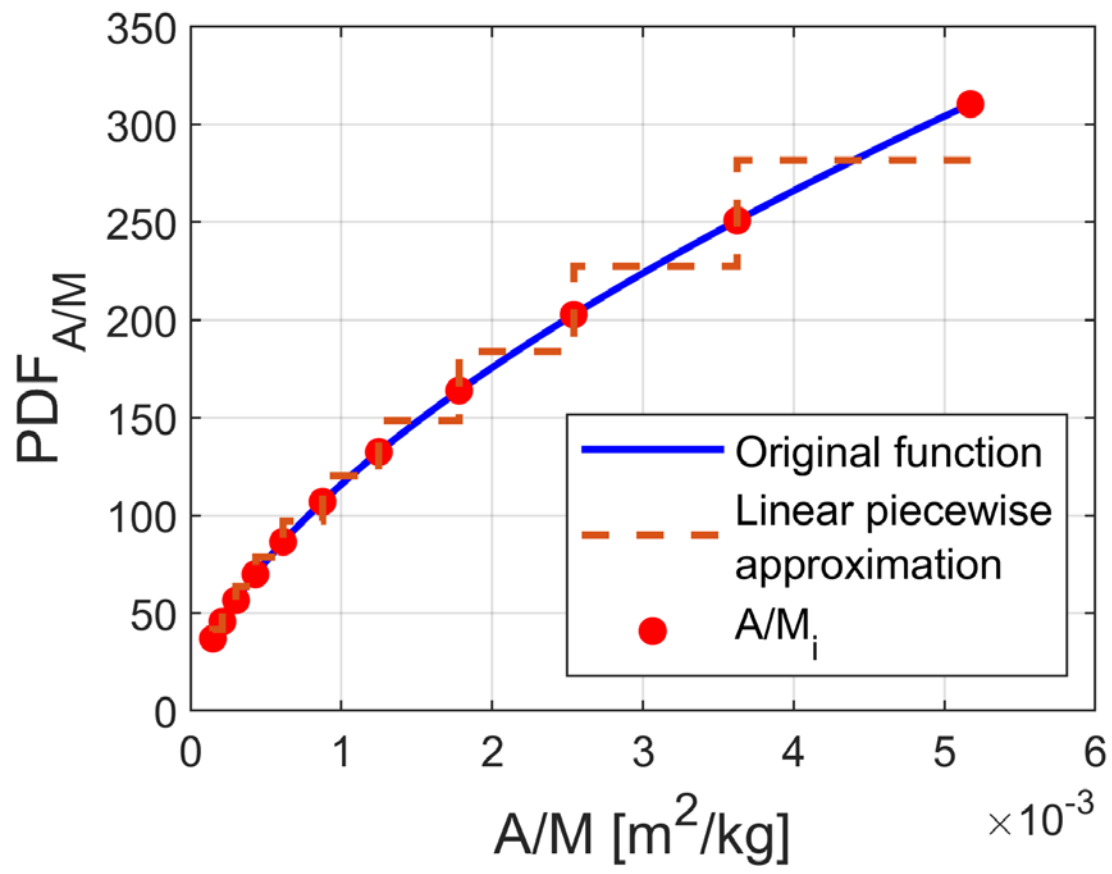

Fig. 9. *A/M* PDF approximation with a piecewise linear function (10 bins).

Each one of the bins is integrated and propagated independently from the others. For each one of them the fragments have a constant size, while the other parameters are assumed to vary following the usual ABM. Using this strategy, the evolution of the domain is greatly simplified: its shape is not elongated anymore but remains compact trough time. Fig. 10 shows the domain evolution in the reduced three-dimensional state space for a 1-meter fragments cloud.

In this case, the Starling fitting can obtain good results for every bin at every time along the trajectory. The use of the binning approximation gives an additional advantage: the range distribution at the time of impact can be approximated as the range distribution at ground, in fact, since the fragments in each bin are identical and the velocity and angle variation are relatively small, the objects move close to each other along the trajectory. As a consequence, at each time the height variation between the fragments is small. This approximation is especially valid for small fragments because of the lower velocity (Limonta et al., 2020). The drawback of this methodology is a reduction of the final accuracy due to the approximation introduced by the discretisation process.

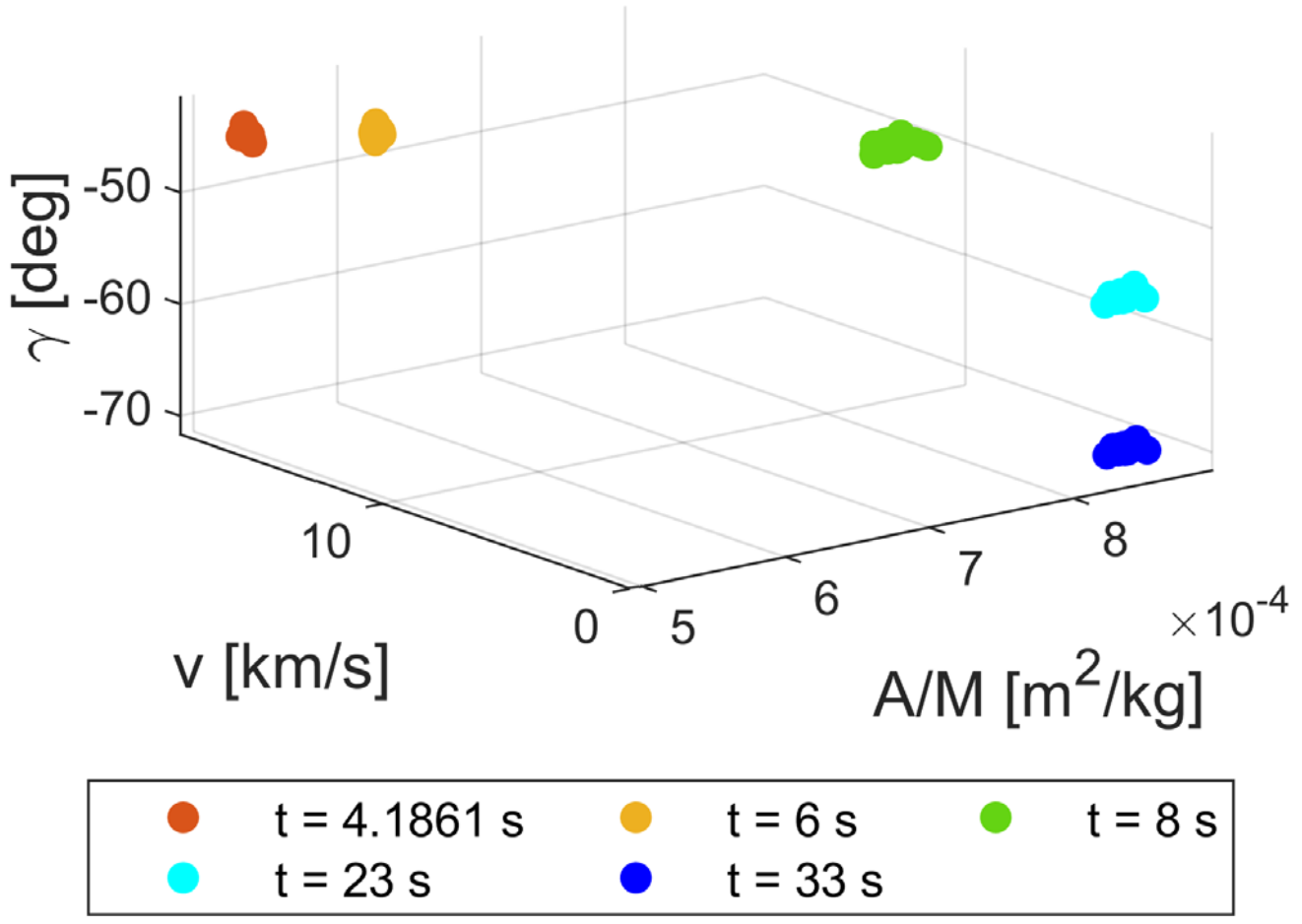

Fig. 10. Domain evolution in time considering a cloud of 1 m fragments.

### *5.5. Footprint estimation*

In this analysis, the focus has been on the meteoroid ground footprint determination and on the fragments' distribution inside this region. These variables are indeed two of the most relevant in a risk assessment analysis. The strewn field is represented by the fragments range distribution at ground. However, the fragments hit the ground at different times while the marginalization is given at a fixed time. Considering a fitting at a time *t*, the joint PDF and its marginals are available. The altitude marginal ($m_h$) is the probability of the fragments to be at a certain altitude at the time *t*, while the two-dimensional marginal of range and altitude ($m_{r,h}$) represents the probability of a fragment to be in a particular position on the trajectory plane at the time *t*. The probability of having a certain range at ground (zero altitude) is then given by:

$$m_{r|h=0} = \frac{m_{r,h}}{m_h} \tag{37}$$

For each snapshot, in which the fragments reach the ground, the marginals have been computed and transformed. To obtain the global range PDF, the marginals are summed and weighted with the normalised number of fragments reaching the ground at every time step as follows:

$$PDF_{range} = \sum_{i=T_0}^{T_f} w_i \, m_{r|h=0_i} \tag{38}$$

where $T_0$ and $T_f$ are the initial and file time of the considered snapshots, and $w_i$ is the weight (normalised number of fragments at ground) of each snapshot. At this point, as outlined in Sec.3.2, the range can be decomposed, by means of the $\chi$ distribution in the $\delta$ and $\lambda$ angular range. Then, the distribution is written in terms of the ground coordinates $X$ and $Y$, obtained by multiplying latitude and longitude by the Earth radius.

$$p_{X,Y} = \frac{m_r \; p_\chi}{R_E} \tag{39}$$

Using this procedure is also possible to obtain a complete description of the evolution of all the other parameters that characterise the fragemnts cloud.

## 6. Monte Carlo validation

In this section, the methodology is applied to recover the ground fragments distribution for a meteoroid entry event. Then the results obtained are compared with a traditional three-dimensional Monte Carlo approach. The test case selected has been obtained considering average parameters among the past meteoroid impacts. Analysing the falls reports, the most common meteorites recorded are ordinary chondrites, belonging to the L or H classes (Carry, 2012). The selected object belong to the L class and its diameter is 5 m. Larger objects are less frequent and there is a chance for them of having a rubble pile structure, which is less consistent with the proposed approach as it assumes a homogeneous meteoroid structure. Smaller bodies, instead, could have a higher strength that prevents the explosive fragmentation. Moreover, the selected size is comparable with the real case scenario analysed in Sec.7. The entry altitude is considered at the Karman line (formal boundary between Earth's atmosphere and outer space), while the entry velocity is taken as an average of several meteoroid entry records (Brown et al., 2011). The flight-path angle is assumed to be 45 degrees, which is considered the most probable entry angle by Shuvalov (Shuvalov and Artemieva, 2002). Starting from the meteoroid size and class, the meteoroid density and all the relevant parameters are inferred as suggested by Cotto-Figueroa (Cotto-Figueroa et al., 2016). Table 1 summarises the selected meteoroid parameters and all the initial entry conditions.

Table 1. Hypothetical entry scenario initial parameters and meteorite characteristics.

| Initial Parameter | Value |
|---|---|
| Meteoroid Class | L |
| Density ($\rho_m$) | $2900\ \mathrm{kg/m^3}$ |
| Diameter ($L_c$) | $5\ \mathrm{m}$ |
| Mass (M) | 189.80 tons |
| Strength (S) | $10^6$ Pa |
| Ablation Coeff. ($c_a$) | $10^{-8}\ \mathrm{s/m^2}$ |
| Drag Coeff. ($c_d$) | 1 |
| | |
| Velocity ($v$) | $17\ \mathrm{km/s}$ |
| Flight Path Angle ($\gamma$) | $-45°$ |
| Altitude (h) | 100 km |
| Longitude ($\lambda$) | 0° |
| Latitude ($\delta$) | 0° |

### *6.1. Results - reduced domain approximation*

Staring from the atmospheric entry (Table 1), the meteoroid states have been propagated with a planar dynamic model (Sec. 3.1) until the fragmentation point (Sec. 2.6). At this point the fragment cloud is generated consisting of 1000 samples, which are propagated towards the ground together with the continuity equation (Sec. 4.1). Also, altitude and range uncertainties have been considered (Sec. 4.2). The samples obtained at ground has been fitted using the GMM approach described in Sec. 5.2 and the ground fragments distribution is obtained as explained in Sec. 5.5. For the case analysed, the falling fragments time windows goes from $t = 9$ s to $t = 36$ s. The time discretisation chosen is of 1 $s$.

For the MC method, the same scheme has been followed, the only difference is the higher number of realisations used and the reduced set of governing equations (i.e., without the density equation). For each realisation, a new meteoroid breakup was simulated, so that $n$ number of fragments, that guarantees the mass conservation, were generated, and propagated to the ground. 500 realisations have been considered with a total for 380000 samples. Fig. 11 represents the PDF at ground obtained with the density-based approach (Fig. 11a) and the comparison with a Monte Carlo simulation (Fig. 11b).

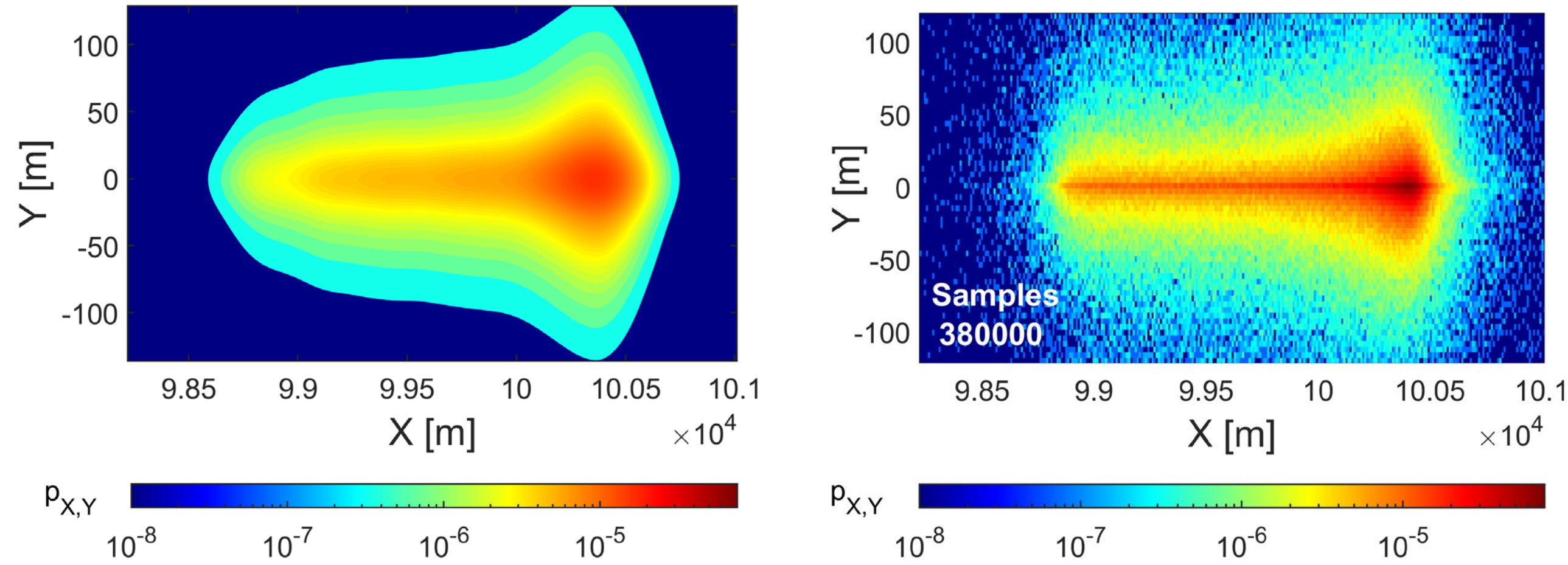

(a) Density-based methodology strewn field. 1000 samples

(b) Monte Carlo estimated strewn field. 380000 samples

Fig. 11 Strewn field distribution obtained with the reduced domain approximation.

The density-based approach provides comparable results with respect to the MC simulation. It is possible to quantify it using the Hellinger distance: it is a metric to measure the difference between two probability distributions. It is the probabilistic analogue of Euclidean distance. Given two discrete probability distributions $p$ and $q$ the Hellinger distance is defined as:

$$D_H = \frac{1}{\sqrt{2}}\sqrt{\sum_{i=1}^{k}(\sqrt{p_i} - \sqrt{q_i})^2} \tag{40}$$

where $D_H$ it is a number between 0 and 1. The closer the value is to 0, the closer the compared distributions are. For the case analysed the Hellinger distance is 0.2935. There is a small difference in the scale of the density: the MC distribution has a higher peak, then it decreases faster than the PDF found with the density-based approach. This is probably due to the approximations used in the three-dimensional methodology and to the time discretisation approximation used for estimating the joint PDF. However, the MC analysis required a very high number of samples to correctly estimate the footprint distribution, while the density-based methodology can reach comparable results with only 1000 samples. Fig. 12 represents the strewn field distribution obtained with a MC analysis using the same number of samples of the density-based approach. In this last case the Hellinger distance is 0.3681. This value is 25% larger with respect to the one obtained with the density-based method. Therefore, with the same 1000 samples, the density-based method gives a better estimation of the strewn field with respect to the MC approach.

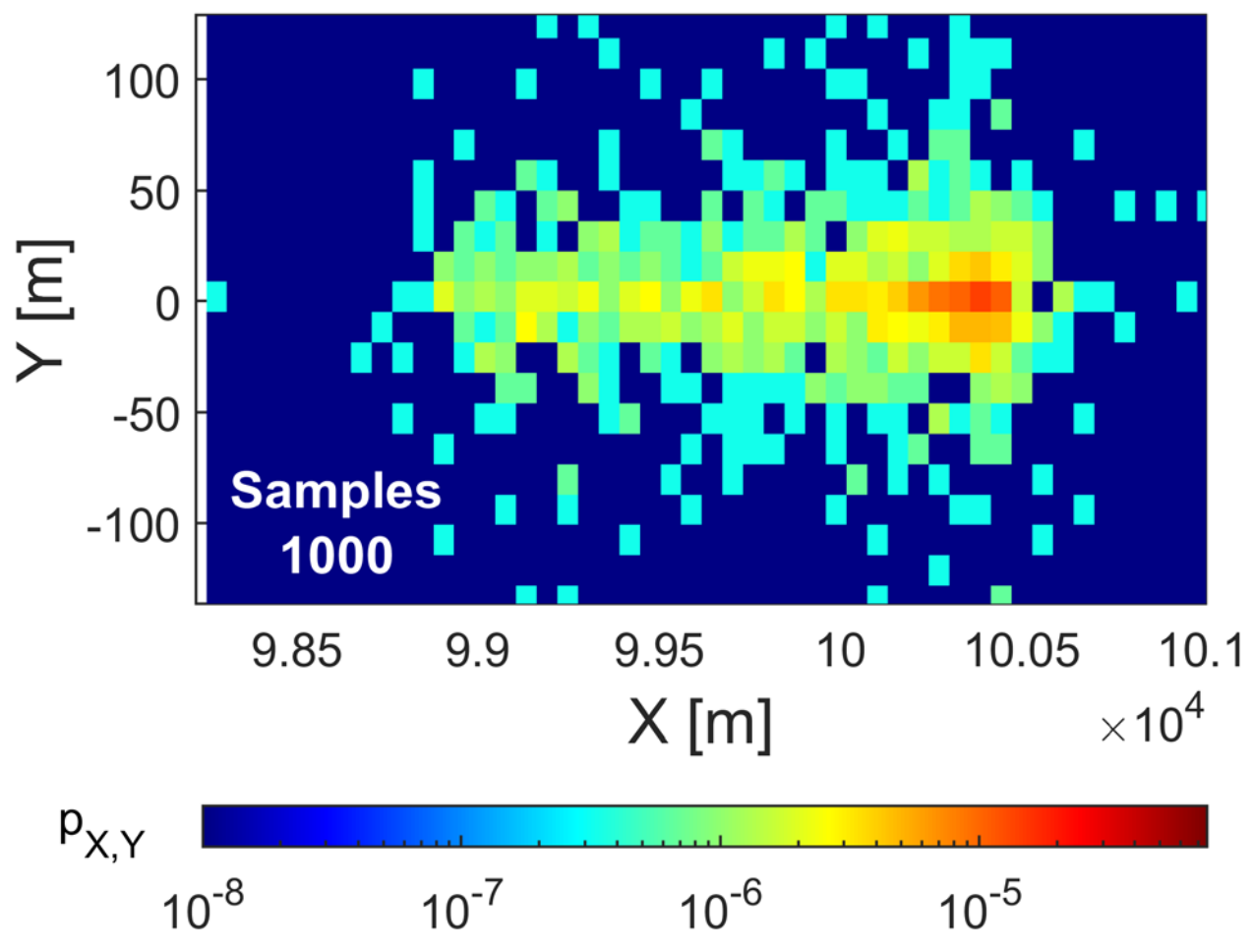

Fig. 12. Strewn field distribution estimated using the Monte Carlo simulation with 1000 samples.

### *6.2. Results – area-to-mass binning approximation*

When using the binning approximation, the steps are equivalent to the reduced domain approximation, but it is possible to exploit one more advantage so that the fitting provides good results for every bin at every instant along the trajectory. The fragments in each bin are identical and the velocity and angle variation are relatively small, each object moves close to the others along the trajectory. Consequently, at each time step, the height variation among the fragments in the cloud is small. Therefore, the range distribution at the time of impact can be approximated as the range distribution at ground. This approximation is especially valid for small fragments because of the lower velocity. Fig. 13 proves the validity of this assumption. It shows the range probability density function fitted at the time of impact and the range distribution at ground estimated using an MC simulation for 1-meter sized fragments. The Hellinger distance between these distributions is 0.1740. It is then possible to approximate the range distribution at the time of impact as the range distribution at ground for 1 m fragments.

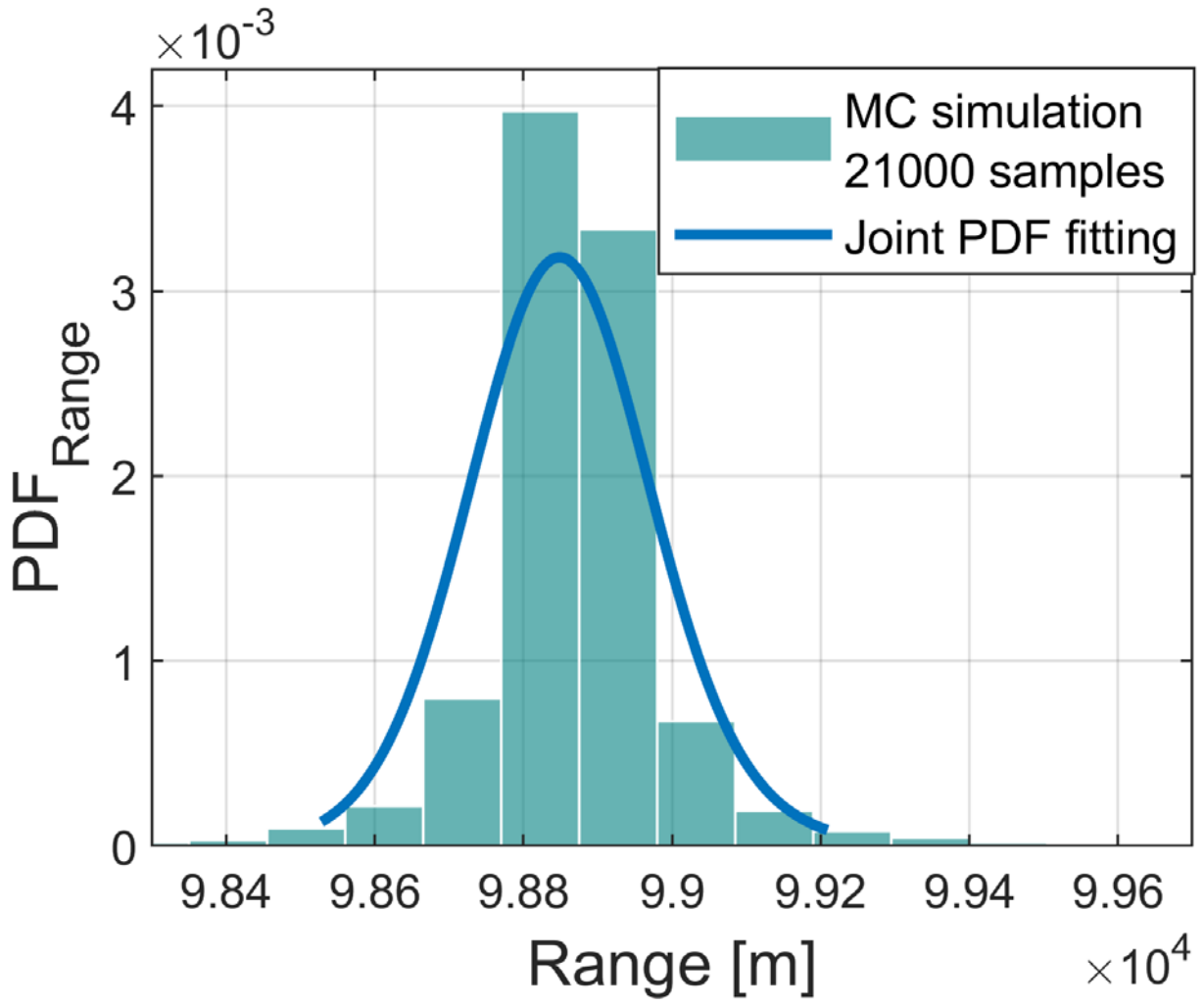

Fig. 13 Comparison in range PDF between the joint PDF fitting and MC simulation for a 1 m fragments bin.

After evaluating the marginal for each bin, the weighted sum is performed (Eq. (38)). However, in the test case considered, the strewn field distributions for each bin are narrow, and their sum does not give a smooth function as shown in Fig. 14. The resulting total range distribution shows peaks in correspondence to each one of the selected bins.

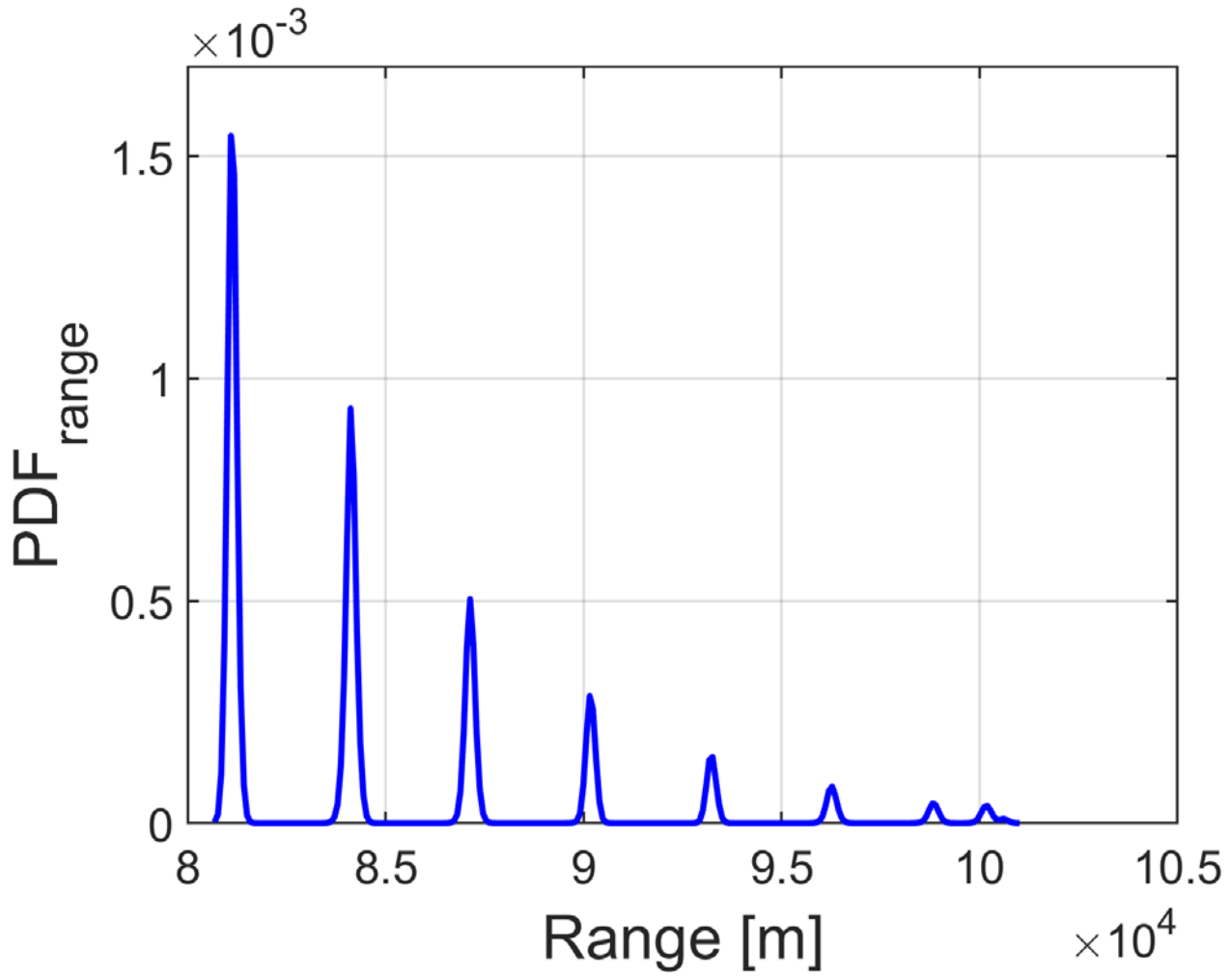

Fig. 14. Weighted sum of all the range PDFs.

This problem may be solved increasing the number of bins; however, because the distributions are very narrow, a much higher number of bins would be required. An alternative strategy, that is the one implemented in this work, is to consider a denser bin grid when sampling the PDF, but to propagate only a small part of the bin. At this point, since the individual range distributions are very narrow, the range PDFs obtained from the propagated bin can be approximated as vertical lines. In this way, the missing information about the non-propagated bins can be recovered simply by interpolating the maximum points of each line (Fig. 15).

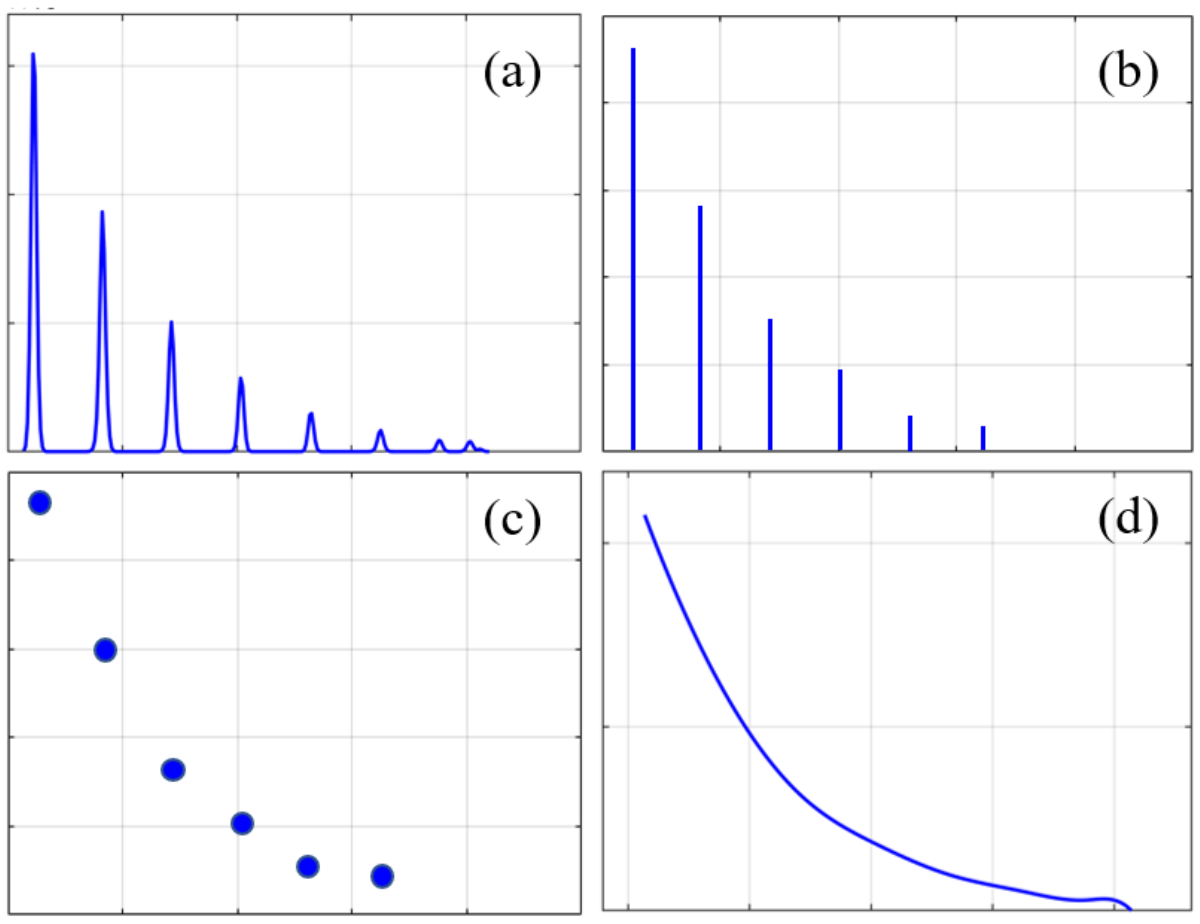

Fig. 16 Scheme used for the interpolation among the propagated range PDFs of the propagated bins.

Following this procedure, we obtain a smooth function that describes the fragments density on ground (Fig. 17). The improved range PDF is obtained considering 64 bins, 10 of which have been propagated and fitted.

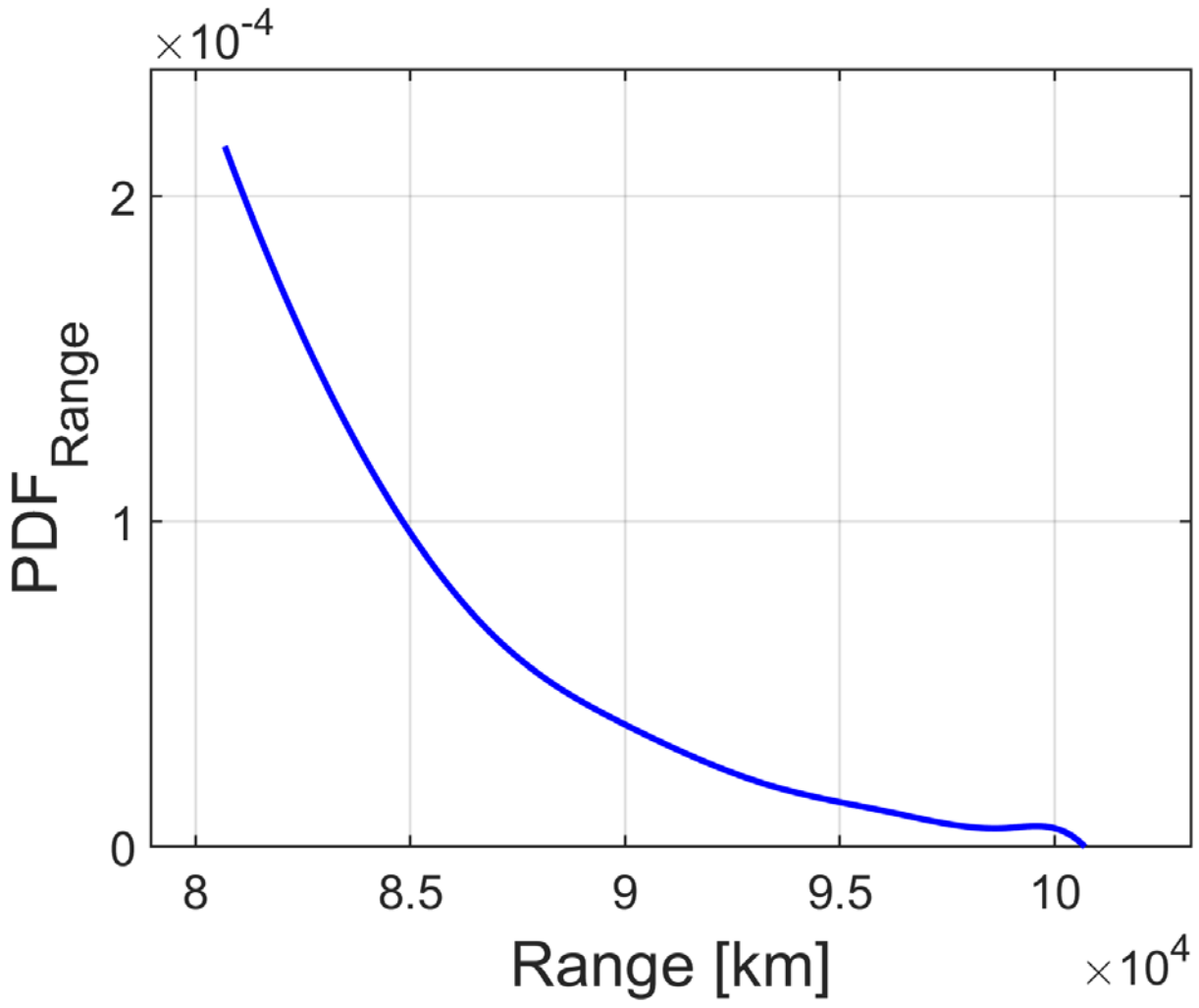

Fig. 17. Mono-dimensional strewn field distribution obtained with the binning approximation.

The function obtained agrees with the MC estimation (Fig. 18) except for in the right end part of the curve. This is probably because the approximation of the time fitting with the altitude fitting assumed in this analysis is weaker for the big and fast fragments. The Hellinger distance between the distributions of Fig. 17 and Fig. 18 is 0.0388.

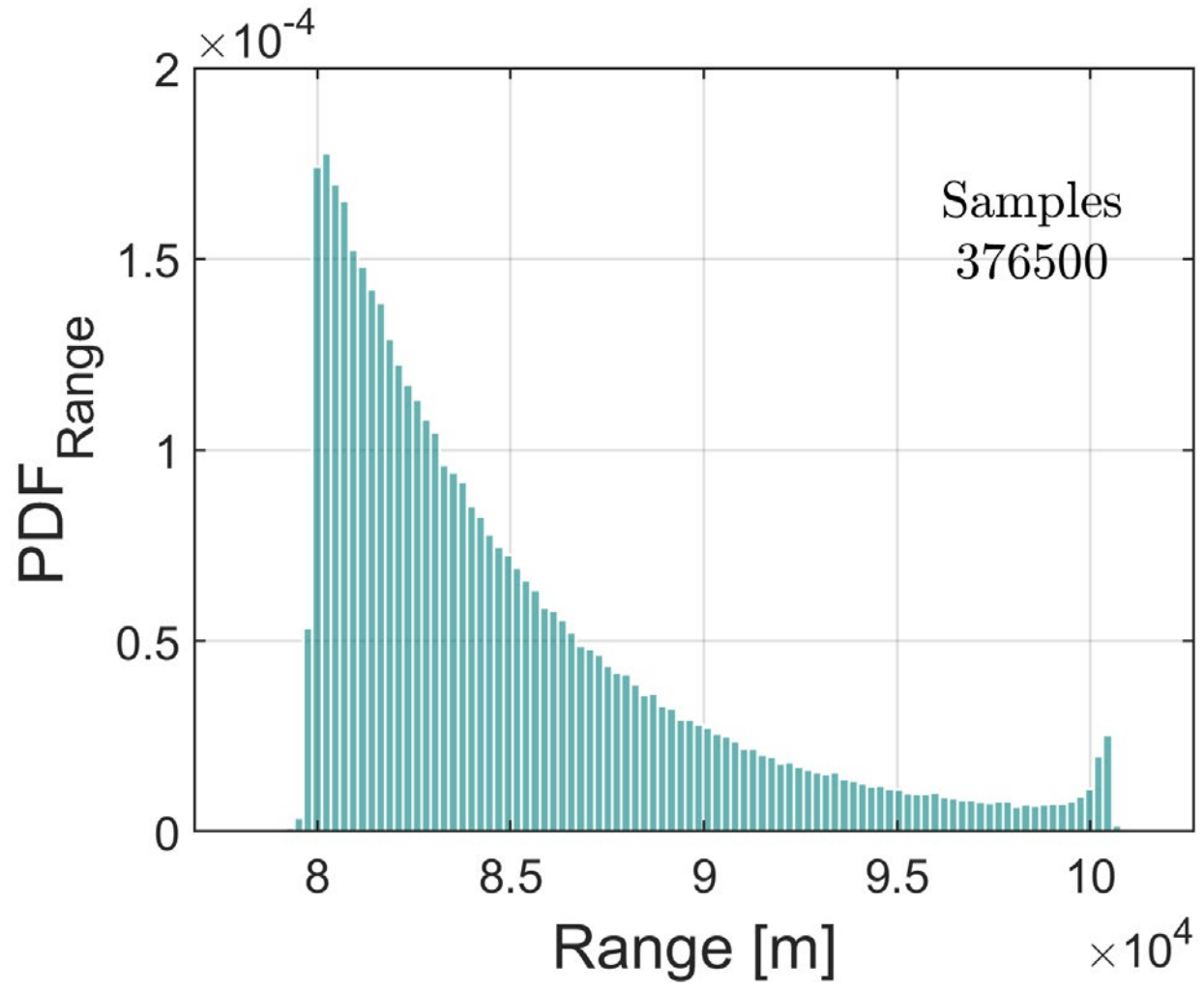

Fig. 18 Mono-dimensional strewn field distribution obtained using the Monte Carlo approach.

Even if the domain remains relatively compact during the propagation (Fig. 10), when analysing the smaller fragments some problems arise in the fitting when $\gamma$ approaches 90 degrees. In fact, the right angle is the accumulation point of $\gamma$: after some time, the lighter bodies will fall vertically causing the collapsing of the flight-path angle dimension. Then, all the other parameters will evolve independently from $\gamma$. This behaviour is not currently supported by the Starling software, so it should be treated with a different approach. When $\gamma \sim 90°$ the range can be considered approximately constant in time; hence, when analysing the smaller fragments, the fitting is performed before the 'vertical fall' event initiates. The smaller the fragments are, the bigger the inaccuracy of this approximation is, because they reach the limit angle at higher altitude. If the trajectory is not exactly perpendicular to the ground, the range distribution evaluated at a high altitude can be different from the one at ground, causing a non-negligible variation in the footprint determination.

The area-to-mass binning approach can be generalised also in the case of a three-dimensional dynamics. In this case, a two-dimensional strewn field distribution (Fig. 19a) is obtained, which can be compared with the traditional MC approach (Fig. 19b).

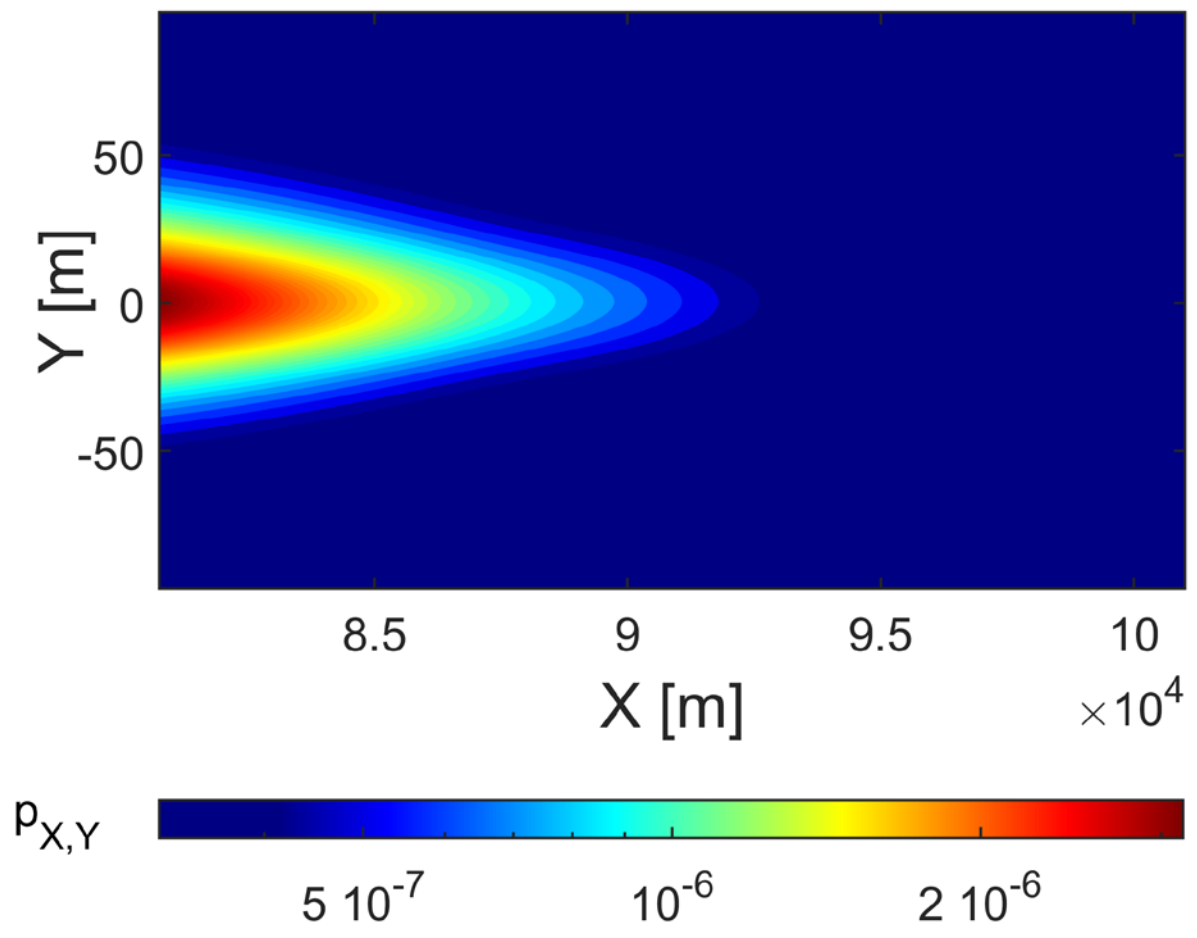

(a) Density-based methodology result. 1000 samples

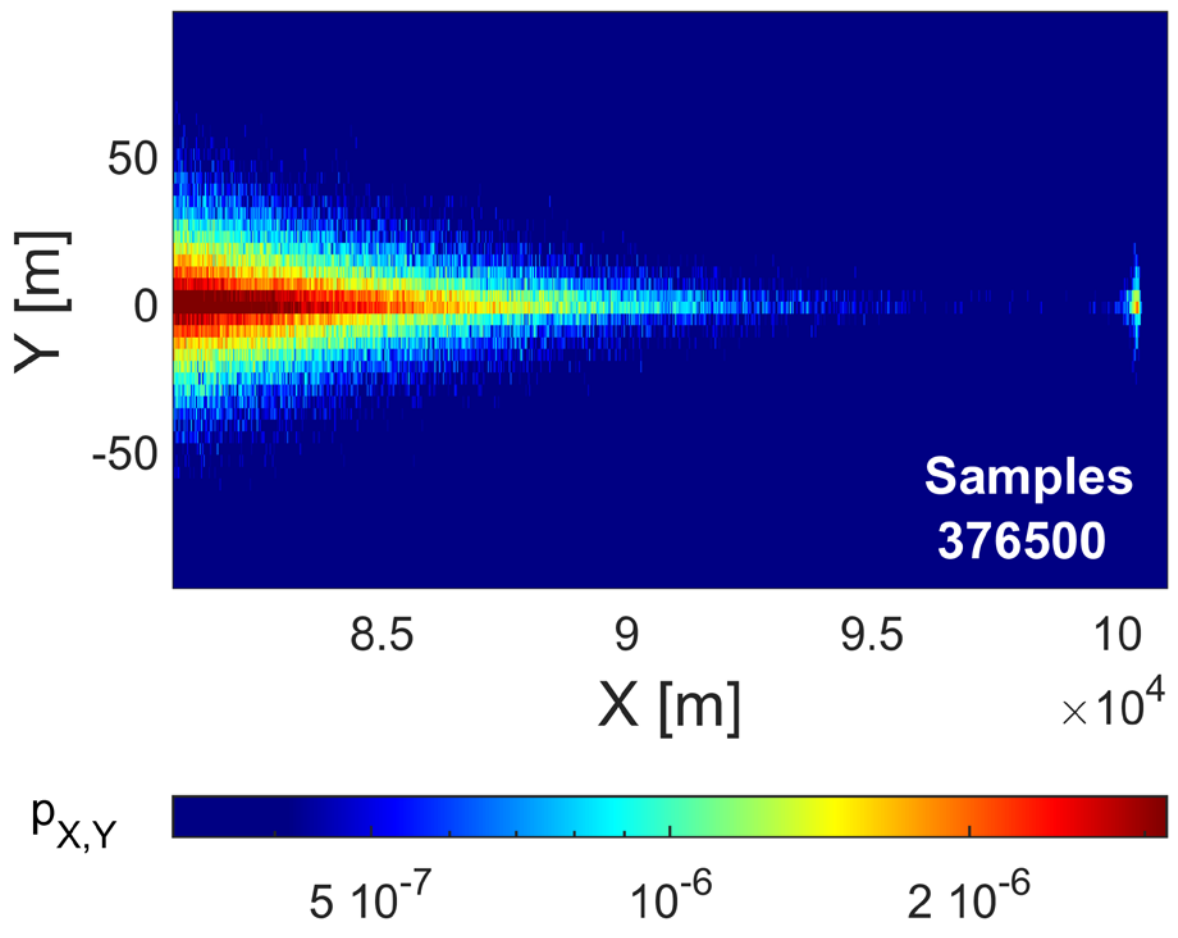

(b) Monte Carlo estimation result. 376500 Samples.

Fig. 19 Strewn field distribution obtained with the binning approximation.

The strewn field obtained with the density-based approach is comparable with the one obtained by the MC estimation. The Hellinger distance between the two distributions is 0.0796. This value is less than one-third of the Hellinger distance obtained by the reduced domain methodology of Sec. 6.1. In fact, in this case, the density magnitude is well estimated and the difference between the density-based and the MC approach is small. The probability magnitude estimated by the MC simulation is slightly higher, that means that the fitting of the density function overestimates the spreading of the fragments outside the symmetry axis. This effect might be related to the constant $\chi$ angle approximation. As in its two-dimension counterpart, also in this case the density-based result, lacks the representation of the second peak downrange. This behaviour is probably related to the binning approximation limitation that does not consider fragments larger than 1 m.

### 6.3. *Results – fragments size distribution at ground impact*

Another variable of interest used for characterising the strewn field is the fragments size distribution ($A/M$) inside the footprint area. This kind of analysis is particularly useful because it relates the position, the size, and the frequency of the fragments on ground.

The binning methodology previously presented is used to evaluate the PDF marginal in a two-dimensional phase space. In particular, the fragment probability density function is considered along the range (i.e., the distance travelled along the orbit ground track) and $A/M$ dimensions.

Repeating the procedure of Sec.6.2, we obtain the $A/M$ vs. range distribution, which is a curved bi-dimensional plane (Fig. 20a) in agreement with the ones estimated with a corresponding MC simulation (Fig. 20b). The smaller fragments have a higher probability density values and are located at the leading edge of the strewn field. The density value decreases progressively as the range increases so that at the opposite side of the strewn field are located the bigger fragments. The Hellinger distance obtained from the comparison between the distributions of Fig. 20 is 0.7781. This value is representative of a lower similarity between the two distributions. The results may also be biased by the extremely localised nature of the distributions, which amplifies the comparison using the Hellinger distance. Additionally, the right part (corresponding to higher area-to-mass ratios) of the MC distribution is not estimated by the density-based approach. A difference can also be observed in the probability magnitude. This behaviour is likely due to the binning approximation used to have a better-behaved domain to perform the fitting.

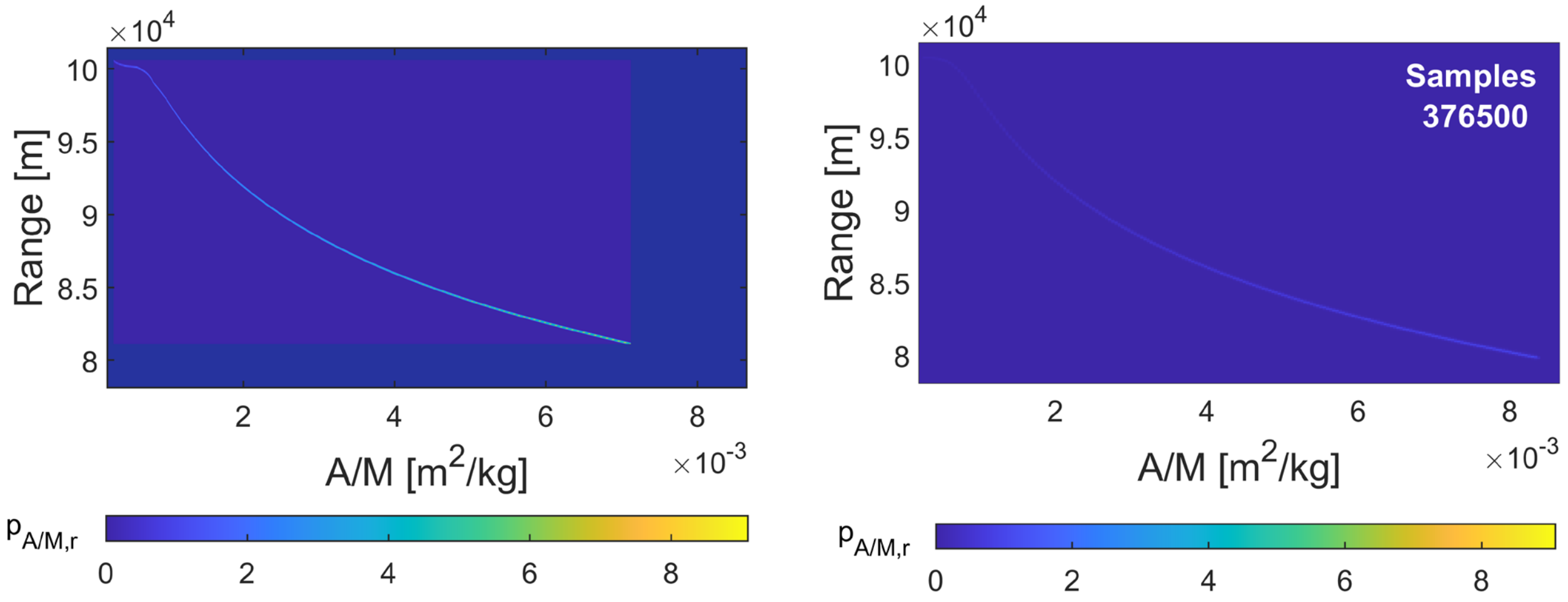

(a) Density-based methodology result. 1000 samples

(b) Monte Carlo estimation result. 376500 Samples.

Fig. 20 Fragments distribution field obtained with the binning approximation.

## 7. Application to the 2008TC3 event

The methodology validated in Sec. 6, is now applied to the analysis of a real entry scenario, the 2008TC3 fall (Farnocchia et al., 2017; Jenniskens et al., 2009; Shaddad et al., 2010). 2008TC3 is an asteroid similar in size to the hypothetical one used in Sec. 6. The purpose of this analysis is to highlight the differences between a real case scenario and a simulated one. From their comparison, we can understand the limitations of the model and gain insight for future developments. Thanks to the high quality of the available data, the analysis of the 2008TC3 fall provides strong and unique observational constraints to test the accuracy of the models used for strewn filed determination.

The asteroid 2008 TC3 was observed on October 6, 2008 and the impact occurred above the Nubian Desert in northern Sudan (Shaddad et al., 2010). The entry velocity relative to the ground was 12 km/s with a flight-path angle γ of 21 degrees (Farnocchia et al., 2017). Jenniskens (Jenniskens et al., 2009) found that the asteroid broke up at an altitude of 37 km. Over 600 meteorites were recovered from the impact site, most of them of small dimension, with a total mass of 10.7 kg (Shaddad et al., 2010). The subsequent analysis of the meteorites indicated that the asteroid was an achondrite and that its original diameter was about 4 m.

Table 2 summarises the parameters of the meteoroid at the entry used in the simulation. However, some of the required data were not directly available from literature. For example, the ablation coefficient and the strength of the meteoroid. In these cases, reasonable assumptions have been made starting from the available information. Regarding the strength, knowing the altitude of fragmentation, it can be reasonably estimated computing the dynamic pressure at that altitude. On the other hand, a direct estimation of the ablation coefficient is not possible from the data available, hence the same value used in Sec. 6 has been adopted (Table 1).

Table 2. 2008 $TC_3$meteorite characteristics and re-entry initial conditions.

| Initial Parameter | Value |
|---|---|
| Meteoroid Class | Urelite |
| Density ($\rho_m$) | $2800\ kg/m^3$ |
| Diameter ($L_c$) | $4\ m$ |
| Mass (M) | $94\ tons$ |
| Strength (S) | $2.2\ 10^6\ Pa$ |
| Ablation Coeff. ($c_a$) | $10^{-8}\ s/m^2$ |
| Drag Coeff. ($c_d$) | 1.8 |
| Velocity (v) | $12.38\ km/s$ |
| Flight Path Angle ($\gamma$) | $-21°$ |
| Altitude (h) | $100\ km$ |
| Longitude ($\lambda$) | 30.54° |
| Latitude ($\delta$) | 21.09° |

It is also important to point out that the drag model used is not an accurate representation of the behaviour of 2008 TC3 in the atmosphere. The asteroid experienced different drag coefficient depending on its orientation, but the constant value of $c_d$ = 1.8 could represent a good approximation as pointed out by Farnocchia et al. (Farnocchia et al., 2017). A final remark is needed on the fragmentation phenomenon: in this simulation the fragmentation is treated as a unique event, but as Jenniskens et al. (Jenniskens et al., 2009) observed, the asteroid showed significant disruption at altitudes around 42, 37 and 33 km. However, the methodology described in this paper currently does not support multiple fragmentation points along the trajectory. For this reason, in the simulation the altitude selected for the fragmentation is 37 km and it is assumed that all the fragments are generated at that instant. This assumption must be taken into consideration when comparing the results of the simulation with the real data. This event was selected for the analysis despite the multiple fragmentation events because of the quality of the data available and because of the insufficient available data on entry events with a single breakup point.

### *7.1. Strewn field*

Fig. 21 shows the ground-projected approach path of the asteroid over the surface of the Earth and the location of the fragments collected after the impact. The masses recovered range from 1.5 g to 283 g and spreads for 29 km along the approach path. The comparison with this scenario is performed using the binning strategy described in Sec.5.4. This was considered the most suitable strategy given it ability to correctly estimates even the smaller fragments distribution, which are predominant for the case in exam. As explained in Sec. 2 the ABM requires as input, together with the states of the asteroid at the breakup, also the maximum and minimum fragments size produced at breakup. Differently from the Sec. 6, in which the size of fragments produced by the ABM are assumed to be between 0.1 m and the 70% of the characteristic length, $L_c$, in this analysis these boundaries have been modified. In fact, the fragments recovered from the meteoroid impact are much smaller than the original size of the body. The smaller fragments size at breakup is assumed to be 5 cm, while the maximum limit is assumed to be 1 m.

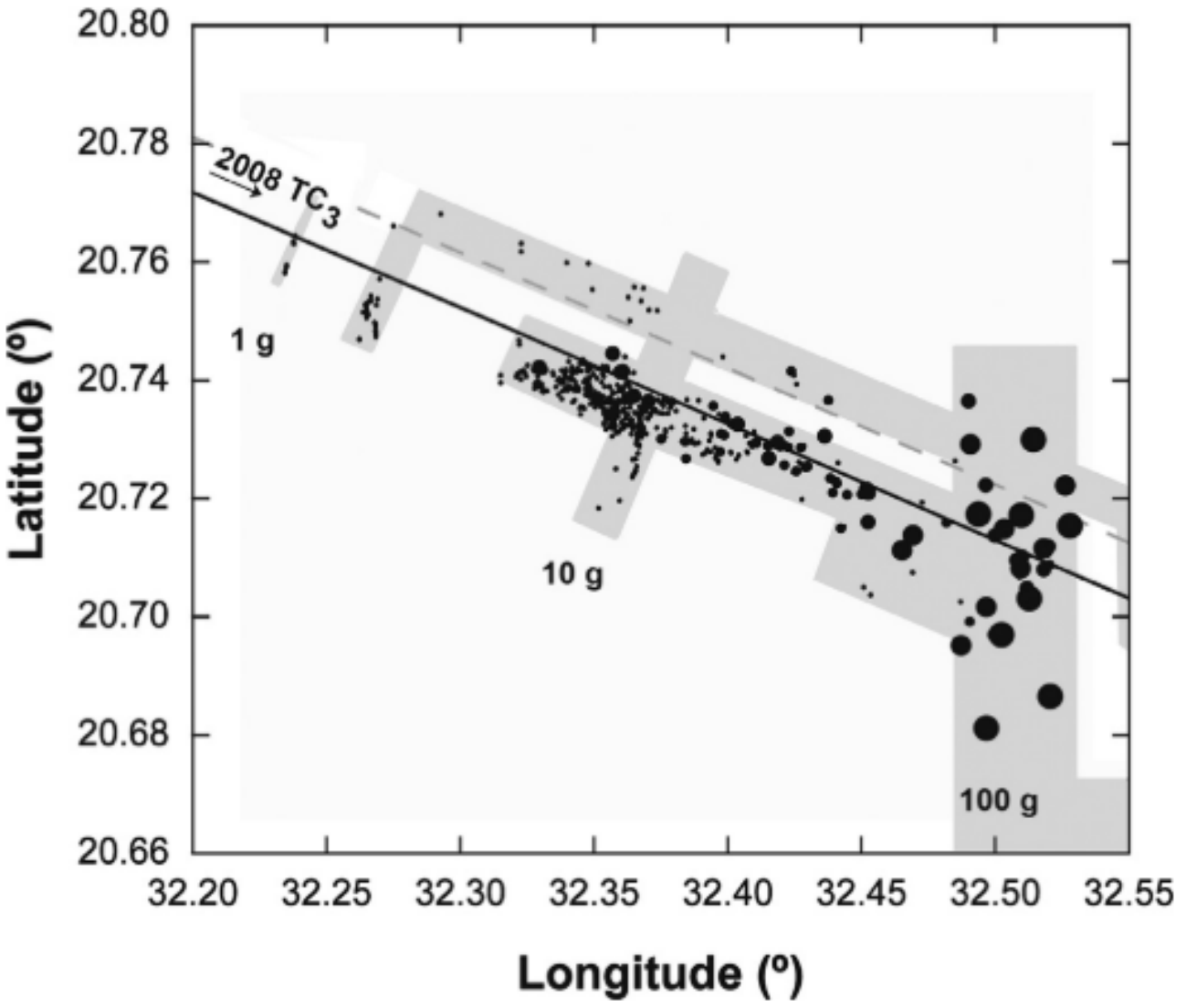

Fig. 21. Drag-free ground track (solid line) and meteorite locations (black dots). Larger dots correspond to larger meteorite sizes. In grey, the area explored by Jenniskens (Jenniskens et al., 2009) and Shaddad et al. (Jenniskens et al., 2009). The dashed line is the ground track used by Jenniskens, while the black ones is the one used by Farnocchia (Farnocchia et al., 2017) that include Earth's J2.

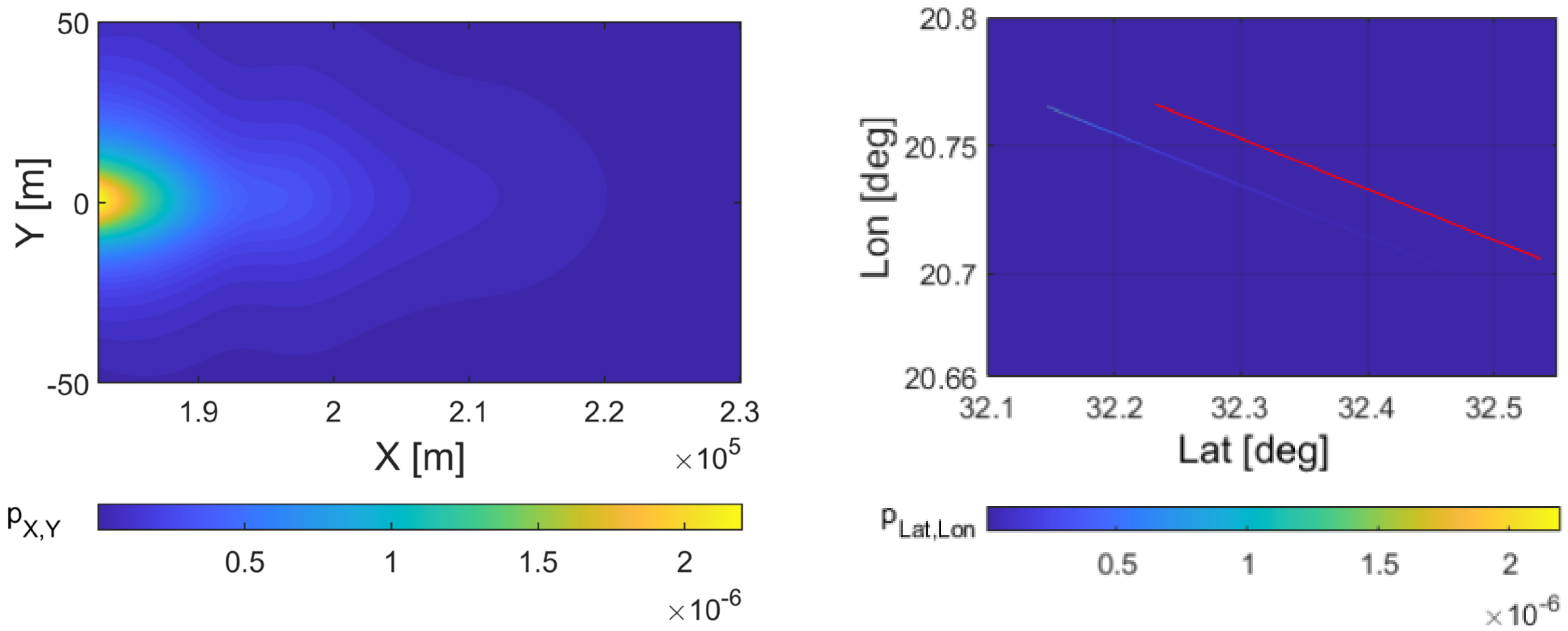

(a) Strewn field along the meteoroid flight track on ground (X axis)

(b) Strewn field in the Fig. 21 reference frame. The coloured 'line' represents the simulation results (Fig. 22a), the red line represents the Drag-free ground track used by Farnocchia et al. (Farnocchia et al., 2017).

Fig. 22 Strewn field shape obtained from the 2008 TC3 entry simulation.

Fig. 22a represents the normalised fragment density on ground: the yellow regions is the one with more concentration of fragments, while the blue region represents the absence of objects. Comparing the strewn field resulted from the simulation with the real one (Fig. 22b), it is possible to observe some differences and some similarities. First, the location of the strewn field on the Earth's surface is reasonably estimated by the model, despite the assumptions made (i.e., non-rotating Earth, constant drag coefficient, constant heading angle, etc.). The comparison shows a good accuracy of the latitude together with a small drift of the longitude towards west. The average distance between the 2 strewn field in Fig. 22b is about 0.02 degrees or 2.23 km.

These results show that the model can reasonably estimate the location of the strewn field. However, the shape of the footprint has notable differences: the predicted footprint can be approximated as a line, while in the real case, even if it is possible to identify a line that interpolates the fragments' location, the fragments are more spread. Additionally, while in the model the line gets narrower as the strewn field evolves toward the left, in the real case the strewn field shows the opposite behaviour (Fig. 21). This behaviour is different also from the typical shapes of the asteroids strewn field, for which the bigger fragments tends to be aligned with the ground track of the asteroid (Norton, 2009). The difference between the model and the real footprint could be caused both by the $\Delta v$ distribution used in the ABM (that has not been modified from the original one in the NASA SBM) and the one fragmentation point approximation. In fact, multiple breakups can contribute to increasing the fragments spread (i.e., the velocity of each fragment is scattered multiple times). An alternative analysis of the entry using a different velocity distribution has been presented in Appendix C.

Another features to highlight in the real strewn field is that the smaller fragments show a south offset with respect to the ground track, which is likely caused by winds at the time of the atmospheric entry. However, winds and side forces are not currently included in the proposed model. Nonetheless, the model correctly estimates the fragments density along the ground track: there is a higher density of fragments at the start of the strewn field (left part of the figure), which decreases going towards the right part of the figure. This behaviour seems different looking at the real strewn field, but it should be noticed that the meteorites have been searched only on the grey rectangles area. Furthermore, it is difficult to identify very small meteorites fragments. For these reasons, it is reasonable to assume that in the left part of the figures more fragments exist, but they have not been found yet. It could also be possible that the smaller fragments were dispersed by cross winds, like dust.

### *7.2. Fragments ground distribution*

Fig. 22 is incomplete, it gives information only on the fragment density, but no information on their size. For this reason, for a more comprehensive analysis, Fig. 22 should be coupled with Fig. 23, which represents the fragments *A/M* distribution as a function of the latitude distance. Fig. 23 shows that the smaller fragments are more and are concentrated at smaller longitudes, while the bigger fragments are rare and concentrated at higher longitudes, coherently with what is observed in the real strewn field.

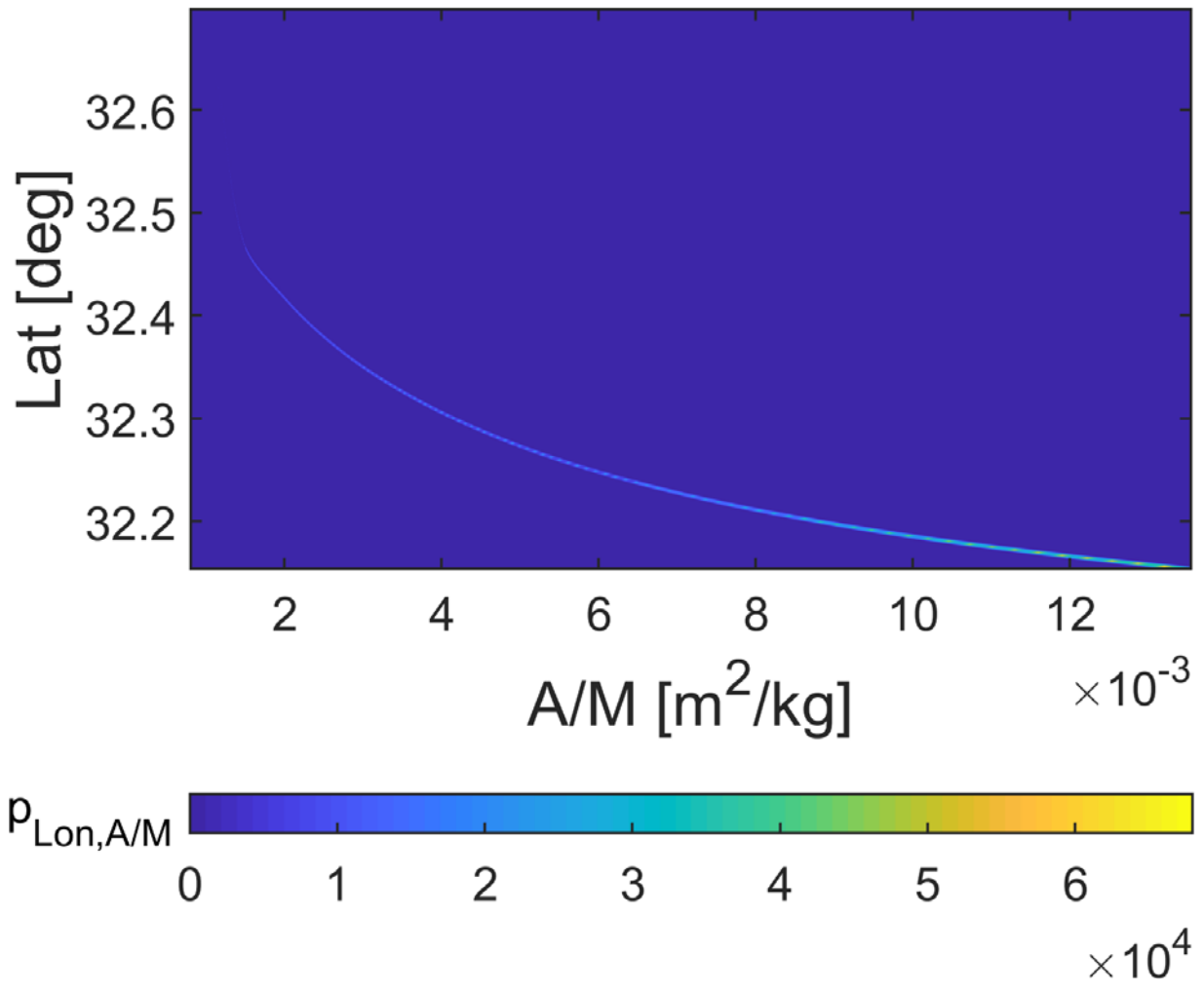

Fig. 23. A/M distribution along the latitude obtained from the simulation.

However, there is a notable difference with respect to the location of the recovered fragments from the 2008TC3 event. The *A/M* range of the simulation is considerably more extended than the one of the observed footprints. An *A/M* value of $12 \times 10^{-3}$ $\mathrm{m}^2/$ kg corresponds to a mass of 130 g, while $2 \times 10^{-3}$ $\mathrm{m}^2/$ kg corresponds to a mass of 28 kg. The presence of the big masses, larger than the meteorites collected on ground, in the simulated strewn field is due to the size limits used in the fragmentation model. An upper bound lower than 1 m was judged infeasible for a fragmentation of a 4 m-asteroid. However, it should be reminded that the proposed methodology is based on a probabilistic model: the absence of big fragments on the real case is compatible with the low probability density value estimated with the continuum approach. However, an overestimation of kilograms sizes fragments could derive from the model uncertainty (i.e., multiple fragmentation points or $\sigma_{ab}$ overstimation). In Fig. 23, the objects of different sizes are more grouped with respect to the real footprint, also in the along-track direction. In fact, the masses below 130 grams are located before the 32.2° of longitude limit, while in the real strewn field, only the masses of 1 gram can be found in the same location. Similarly, to the lack of spread in the direction perpendicular to the along track direction, this behaviour might be due to the underestimation of the $\Delta v$ at the breakup or to the multiple fragmentation points experienced by the real event.

## 8. Conclusion

The objective of this work is to propose new methodology to model the fragmentation of large meteoroids and small asteroids during entry events. This approach is based on the statistical description of the fragmentation event, which provides the distribution of the characteristics of all the fragments generated at the breakup. The model describes the fragments cloud by means of a continuous distribution function defined in the *A/M*, $v$, $\gamma$ space. To take advantage of the continuum formulation of the fragments cloud, a density-based propagation methodology has been exploited to predict the evolution of the fragments along the entry trajectory until ground impact.

From the presented analysis, the distribution of the meteoroid fragments during the re-entry is characterised by a peculiar domain, which has an elongated shape, evolves quickly in time, and presents some challenges for the fitting routines. Simplifying the domain and focusing the analysis on the larger fragments have been proven feasible strategies for the prediction for the fragments strewn filed with the density-based method, with results comparable to Monte Carlo simulations. This is a relevant result, because the main drawback of the Monte Carlo simulations is the high number of samples required to provide a good estimate of the distributions. The density-based methodology, instead, reached comparable results with a limited number of samples.

The results obtained studying the 2008TC3 impact showed a good approximation of the landing site. However, differently form the real strewn field, the predicted one is narrower both in the along-track and cross-track direction. Furthermore, the location of the fragments inside the strewn field does not match accurately the one recorded on ground. This reduced accuracy can either derive from the formulation of the ABM, which underestimates the velocity difference between the fragments after the breakup, or from the approximation used in the dynamics that does not account for the Earth's rotation, the presence of crosswinds, and the occurrence of multiple breakups, whose contribution can be significant for smaller fragments. Probably each one of the previous points contributed in part to the differences between the two strewn filed.

At the current state of development, the model can predict the single fragmentation event and gives an estimation of the ground footprint and the fragments distribution inside it. However, a more refined breakup model may be required, which is directly derived from asteroid breakup data, instead than borrowed from satellite breakups. Furthermore, in some cases the generated fragments may experience further fragmentations during the descent. The model used in this work assumes only one breakup point along the meteoroid trajectory. A further development, which will improve the model accuracy and flexibility, can consider the presence of multiple fragmentation events. This model may also be integrated in others existing parent-child models to allow for a statistical representation of the cloud of small fragments. In fact, the structure of the methodology presented in this paper allow the integration of more representative fragmentation models, when available, without any modification on its other parts.

## Acknowledgements

This project has received funding from the European Research Council (ERC) under the European Union's Horizon 2020 research and innovation programme (grant agreement No 679086 - COMPASS).

## Appendix A. Probability density function transformation

By definition, a given probability density function, $p_x = p(x)$ is required to sum to unity if integrated over the whole state space $x.$ The fragment distribution, instead, is described by the phase space density $n_x$, that is the number of fragments in an infinitesimal volume around a state $x$. The integration of it over the full domain yields the total number of fragments, *N*.

$$\int_{-\infty}^{+\infty} n_x \, dx = N \tag{41}$$

Since $n_x$ and the $p_x$ only differ via the normalization constant, the space density function can be treated as a probability density function.

At this point, considering a one-to-one change of variables, $y = \boldsymbol{\varphi}(\boldsymbol{x})$ with $\boldsymbol{x}, \boldsymbol{y} \in \mathbb{R}^n$ . If $\boldsymbol{\varphi}$ is differentiable, then the probability density function $p_y$ can be derived from the probability density function $p_x$ in the following way (Ghahramani, 2015):

$$p_y(\boldsymbol{y}) = \frac{p_x\left(\boldsymbol{\varphi}^{-1}(\boldsymbol{y})\right)}{|\det \boldsymbol{J}|} \tag{42}$$

With the Jacobian $\boldsymbol{J} \in \mathbb{R}$^$(n \times n)$ defined as

$$J_{i,j} = \frac{\partial \varphi_i}{\partial x_i} \tag{43}$$

If the function $\boldsymbol{\varphi}$ is not invertible, the probability then is the sum of all the possible inputs (Frey and Colombo, 2020).

## Appendix B. Characteristic length bounds analysis

As mentioned in Sec. 2.1, when different characteristic length bounds are considered, the strewn field may change, depending on which fragments size are neglected. Fig. 24 shows the comparison of different entry simulations considering different characteristic length bounds. These simulations provide a qualitative comparison of the strewn field shape and show the difference in the number of generated fragments for varying characteristic length bounds. The meteoroid considered for the simulation is the same test meteoroid analyzed in Sec. 6. Depending on the different $L_c$ bounds the ABM generate a different number of fragments so that the total meteoroid mass will be conserved. As expected, the larger the fragments the fewer the number of fragments generated. Fig. 24 shows that the strewn field shape changes substantially only when the analysis is limited to only small fragments (upper left) or to large fragments (lower right). In the first case, the number of fragments is high but is distributed only over a small area, in the second case the number of fragments is small, and they are concentrated along the meteoroid trajectory. The other two cases (upper right and lower left) in which the $L_c$ bounds are wider, the fragments on ground are more distributed and the strewn field shape and range is relatively unchanged.

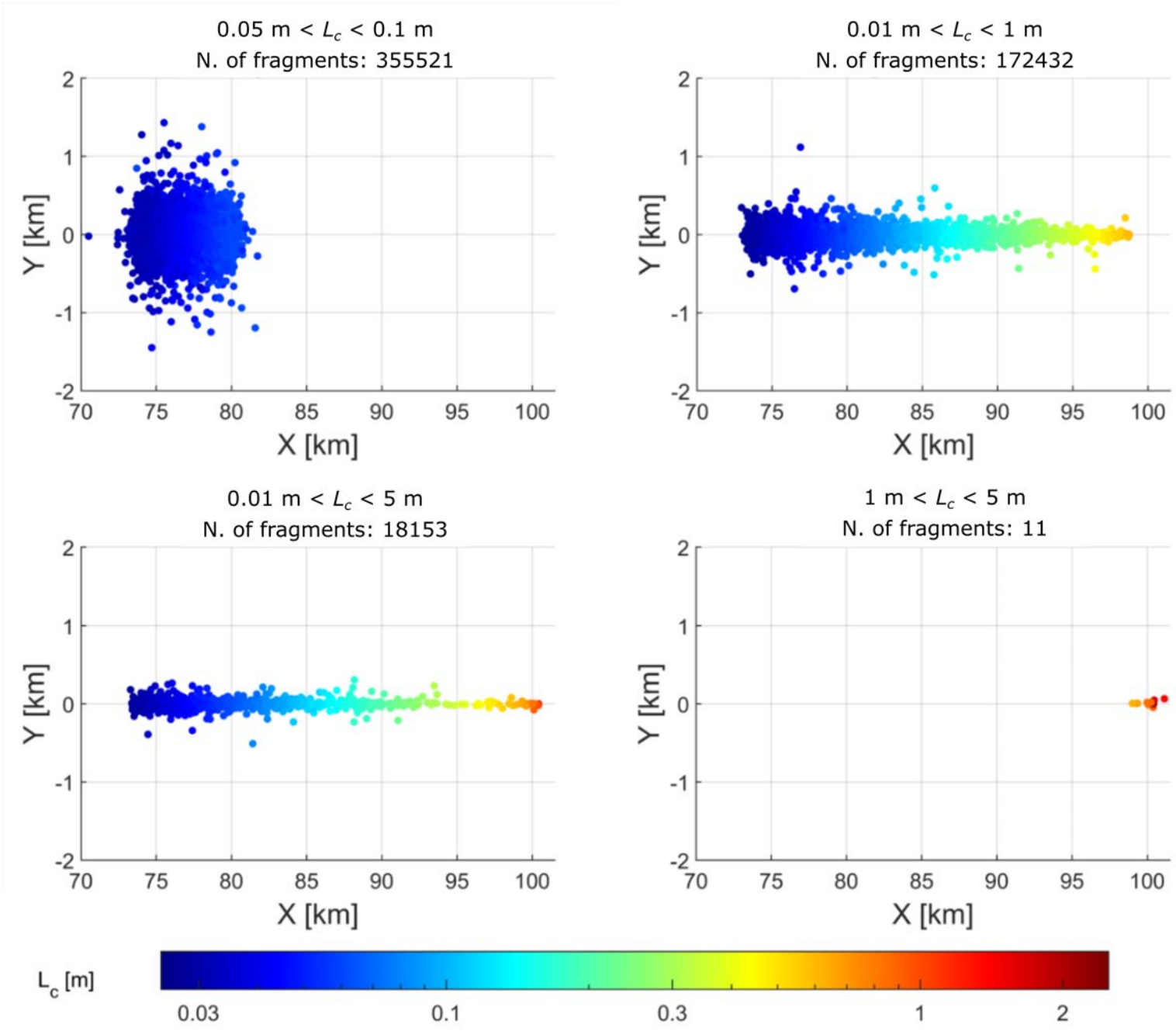

Fig. 24 Strewn field obtained considering different $L_c$ bounds for the Test meteoroid entry (Table 1). Each point represents a child fragment and its colour its diameter. The bigger the fragments the larger is their range on ground.

## Appendix C. Alternative velocity distribution

The results obtained with the application of the ABM to a real case scenario suggest that the velocity distribution is critical for the fragments cloud evolution and for the subsequent strewn filed formation. In this section the 2008TC3 entry will be analysed using a traditional velocity distribution that will substitute the one derived from the NASA SBM. As pointed out by Register (Register et al., 2020) the most commonly adopted velocity model to describe meteoroid fragmentations derives from the research of Passey and Melosh (Passey and Melosh, 1980). They calculated the dispersion velocity by which two spherical fragments separate after undergoing a breakup event:

$$v_{disp} = \left(3\,C\,\frac{D_1}{D_2}\frac{\rho}{\rho_m}\right)^{1/2} v \tag{44}$$

Where $v_{disp}$ is the velocity increment in the perpendicular direction. $D_1$ and $D_2$ are the parent and child object diameters, respectively, $\rho$ is the air density and $\rho_m$ is the meteoroid density, $v$ is the airstream velocity, i.e., the meteoroid velocity, and $C$ is the minimum edge-to-edge separation distance (in multiples of $D_1$) for the fragments to have independent bow shocks. The authors estimated the value of $C$ to be on the order of 0.5 (Passey and Melosh, 1980).

It has been demonstrated that for two identical fragments the dispersion velocity becomes constant for separations greater than $C$ and its value is three times smaller than the value predicted by Eq.1 (Register et al., 2020). This result agrees with findings from McMullan and Wheeler (McMullan and Collins, 2019; Wheeler and Mathias, 2019). In this analysis, the 2008TC3 entry has been simulated using both the original Passey and Melosh velocity distribution (P&M) and considering the velocity reduction factor demonstrated by Register (Register et al., 2020). In this second case the velocity correction, proposed for two identical fragments, is extended to the whole debris cloud. The expression of the dispersion velocity including the considerations by Register is therefore:

$$v_{disp,R} = \mathrm{f_R}\, v_{disp} \tag{45}$$

where the correction factor $\mathrm{f_R} = 0.34$. It is useful to compare the velocity magnitude increment given to each fragment for each of the proposed model and with the NASA SBM (Fig. 25). The NASA SBM is probabilistic while the alternative models are only a function of the fragments size. Given its statistical nature, the NASA SBM can assign high velocity increments event to larger fragments, even though these types of events have very low probability of occurrence. On the contrary, the Register and P&M models do not allow for this random variation and admit a maximum velocity threshold, depending on the characteristic length cut off. The Register model for example admits a $\Delta v_{max} =$ 50 m/s for 10 cm fragments. Fig. 25 shows that, on average, the velocity estimated for fragments of the same size with the NASA SBM is lower than the one obtained by the Passey-Melosh and Register models; however, the presence of a long-tailed distribution such as the log-normal can introduce outliers with high $\Delta v$ compared to the size of the fragment.

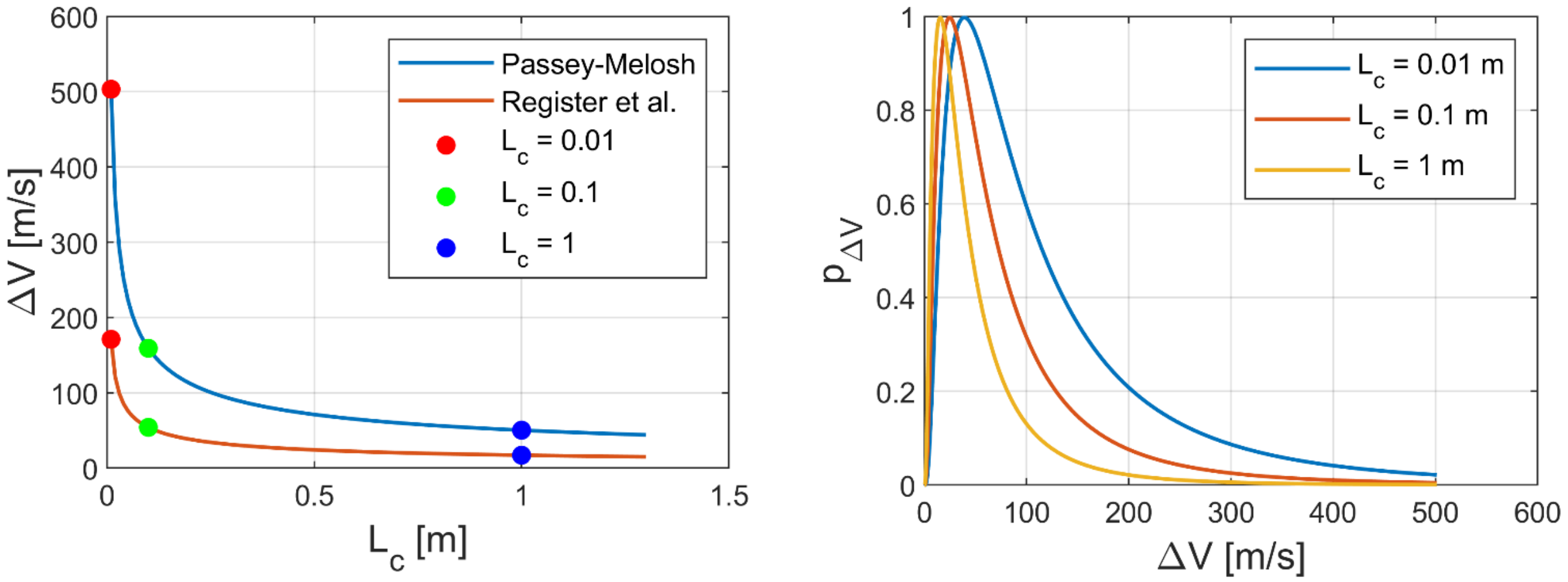

(a) P&M and Register velocity model. Curves have been represented considering different characteristic length values.

(b) NASA SBM velocity increment magnitude probability. Different curves have been represented depending on different fragments characteristic length.

Fig. 25 Velocity magnitude increment given to each fragment during the fragmentation process.

For each one of the velocity models discussed, a MC analysis has been performed evaluating the meteoroid fragmentation and the ground strewn field. Results are summarized in Fig. 26. Fig. 26a, Fig. 26c, Fig. 26e represent a two-dimensional binning of the fragment cloud generated by the ABM. The fragments have been grouped depending on their size and velocity increment. On average, the fragments velocity increment generated by the NASA SBM are lower than the ones obtained with the P&M model and comparable with the one obtained with the register model. The only difference is that the NASA SBM velocity distribution can generate a small amount of high velocity fragments, which may not be physically consistent with the with failure of a meteoritic material subject to an external pressure. However, looking at the strewn field obtained in each case (Fig. 26b, Fig. 26d, Fig. 26f) there is only a small difference. It should be noted that in Fig. 26b and Fig. 26d the domain in the Y direction has been scaled in a way that it was easier to compare the different strewnfield. The original extension of the domain is bigger, especially in Fig. 26b, due to the presence of sporadic fragments with high velocities. However the probability associated with these fragments is negligible with respect to the order of magnitude of the strewn field. The main part of the strewn field in in agreement considering all the velocity models and also the velocity magnitude increment has the same order of magnitude.

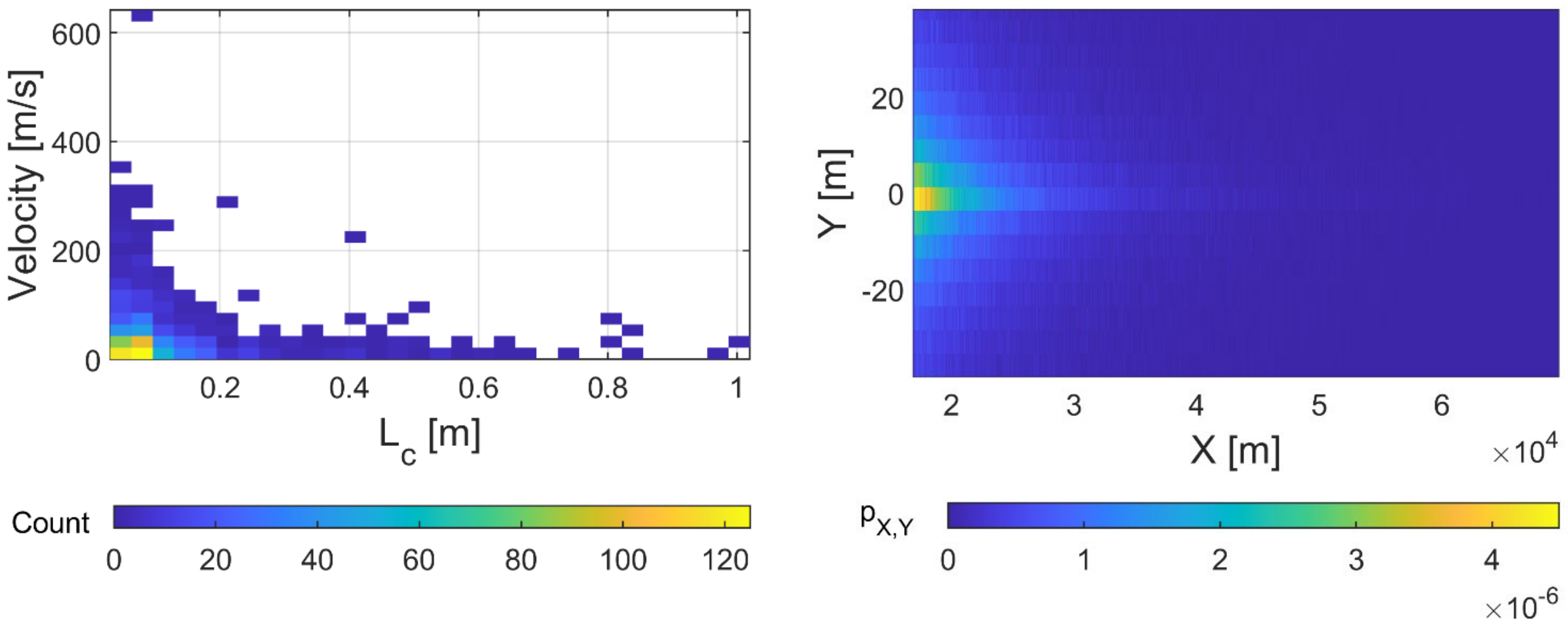

(a) Fragments generated using the NASA SBM velocity distribution. 6371 fragments

(b) Strewn field obtained using the NASA SBM velocity distribution. 317550 samples. The results have been scaled so that it was possible to compare with other distribution results.

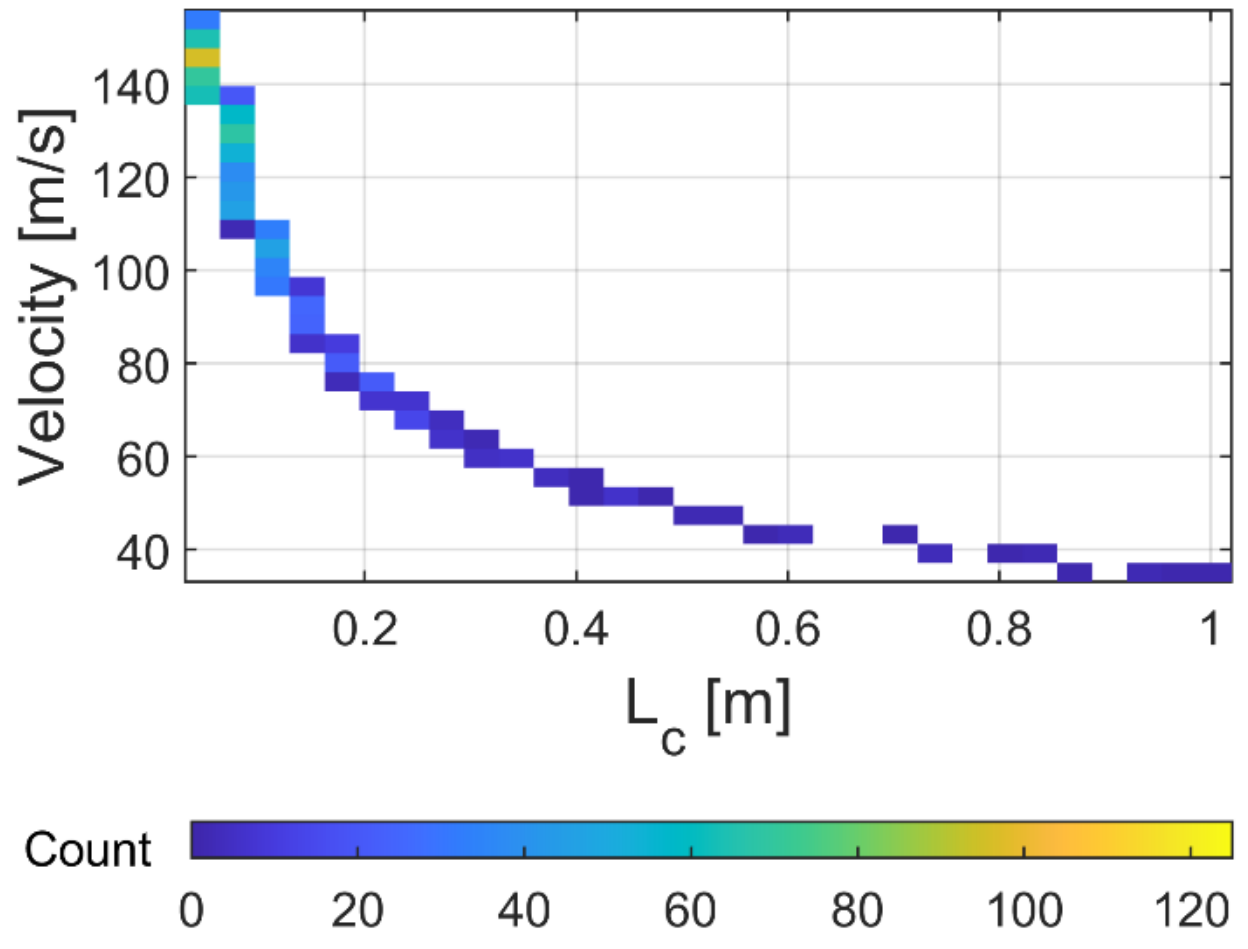

(c) Fragments generated using the P&M velocity distribution. 6371 fragments

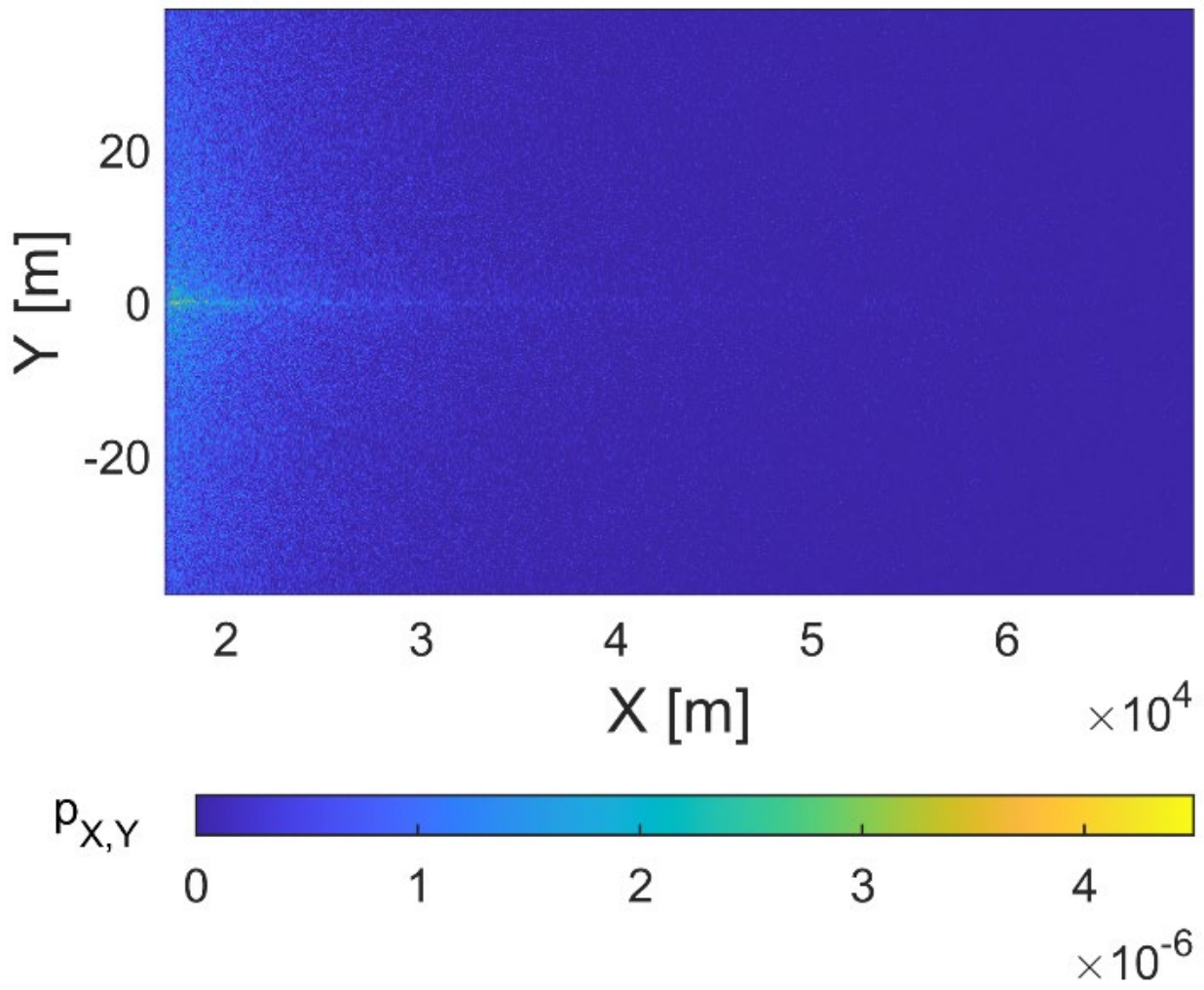

(d) Strewn field obtained using the P&M velocity distribution. 317550 samples. The results have been scaled so that it was possible to compare with other distribution results.

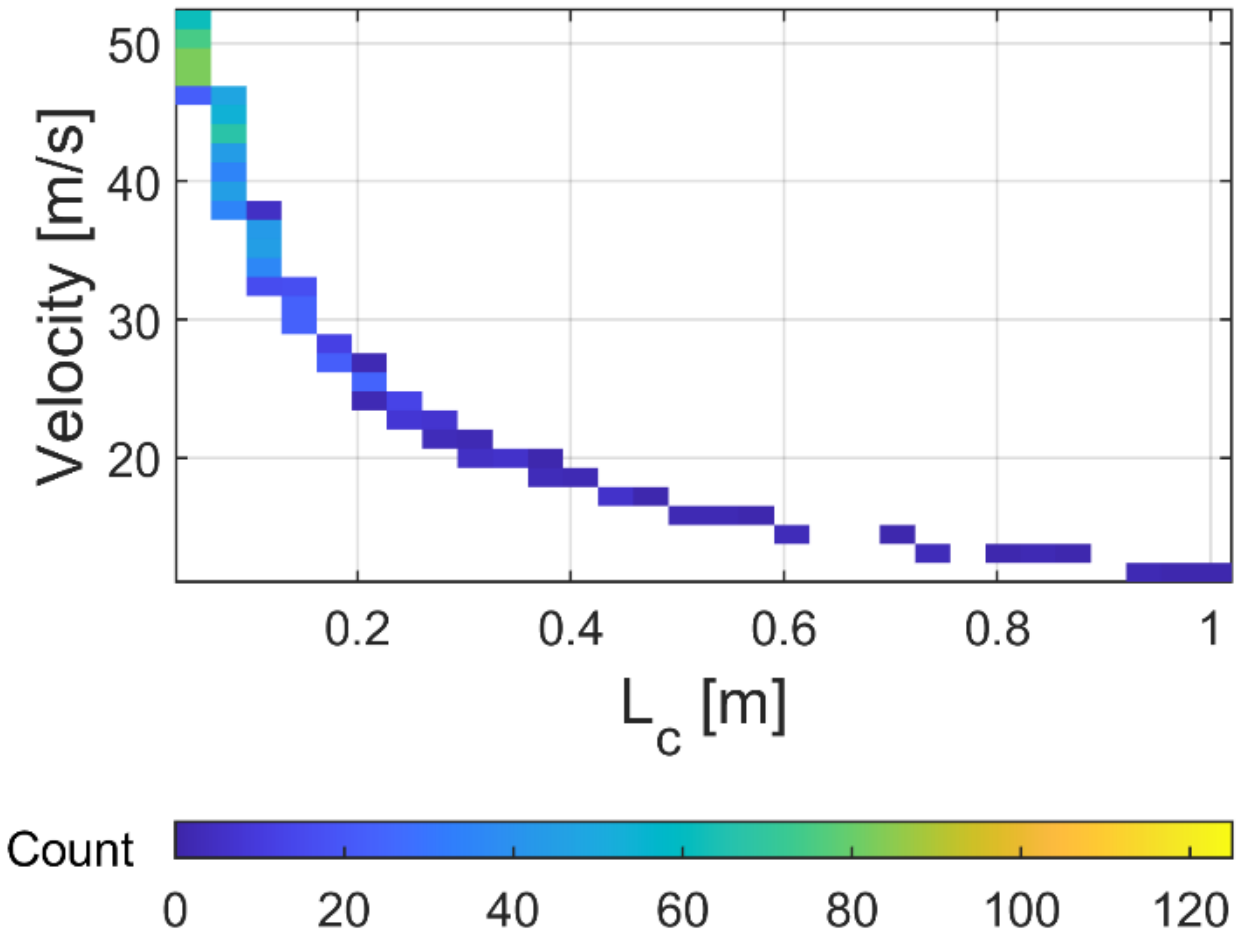

(e) Fragments generated using the Register velocity distribution. 6371 fragments

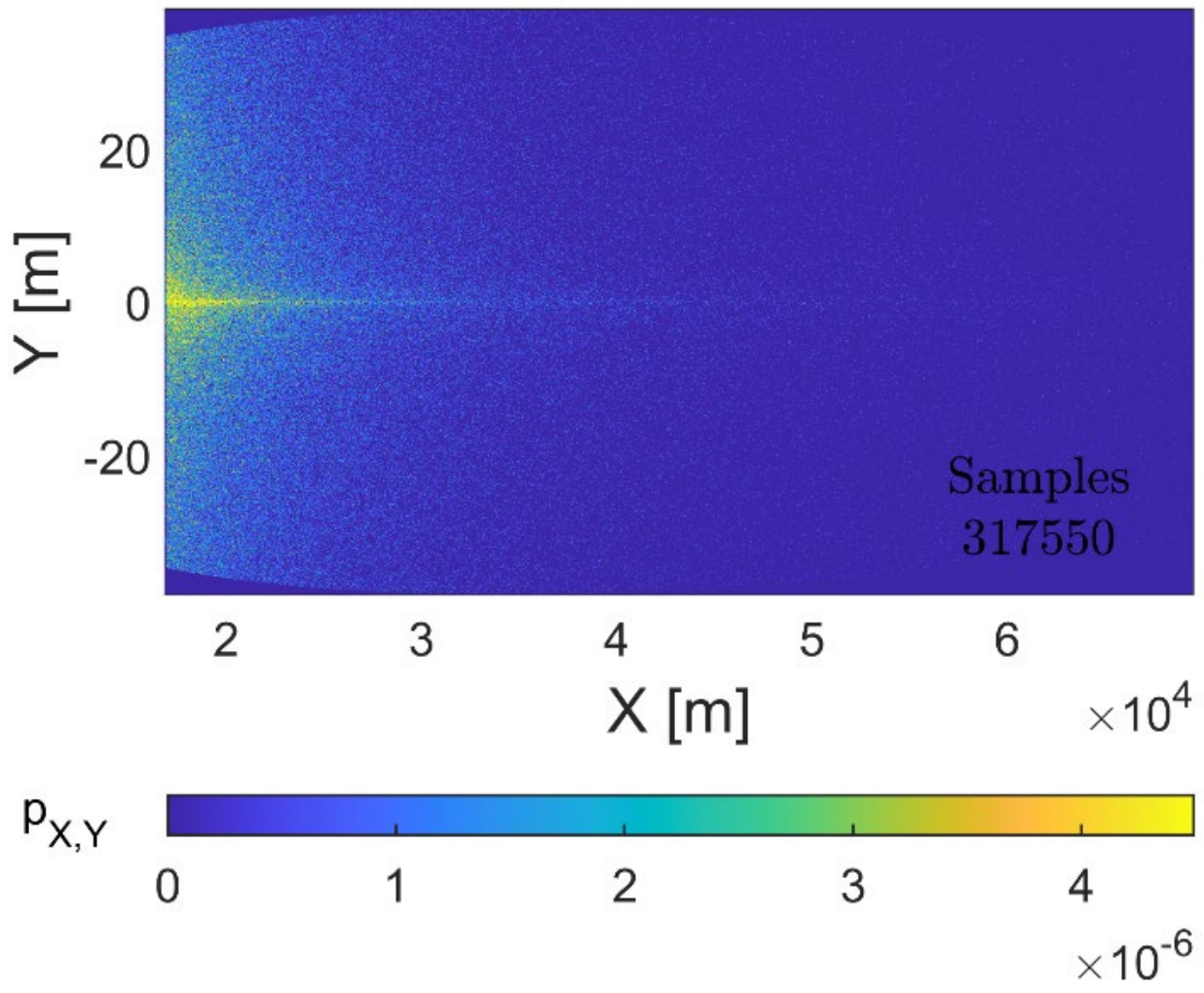

(f) Strewn field obtained using the Register velocity distribution. 317550 samples.

Fig. 26 2008TC3 entry simulation results considering different velocity models.

## Appendix D. Nomenclature

| | |
|---|---|
| $L_c$ | Characteristic length |
| $N_c$ | Number of fragments generated larger than a given $Lc$ |
| $N_{tot}$ | Total number of fragments |
| $c_d$ | Drag coefficient |
| $p_{a,b}$ | Probability density function in a,b space |
| $\rho_m$ | Meteoroid density |
| $\sigma_{ab}$ | Ablation coefficient |
| $h$ | Altitude |
| $A/M$ | Area-to-mass ratio |
| $M$ | Meteoroid mass |
| $R$ | Meteoroid radius |
| $St$ | Meteoroid strength limit |
| $f$ | Power factor of the $Lc$ distribution |
| $k$ | Tuning parameter for mass conservation |
| $n$ | State space density |
| $v$ | Velocity |
| $w$ | Bin relative weight |
| $\mathcal{N}$ | Normal distribution |
| $\gamma$ | Flight path angle |
| $\delta$ | Latitude |
| $\zeta$ | Angular range |
| $\lambda$ | Longitude |
| $\rho$ | Air density |